\documentclass{SciPost}

\usepackage{amsmath,amssymb,mathtools}
\usepackage{mathrsfs,bbm}
\usepackage{cleveref}
\usepackage{subcaption}
\usepackage[toc,page]{appendix}
\usepackage{graphicx}
\usepackage{calc}
\usepackage{listings}
\usepackage{dsfont}
\usepackage{stackrel}
\usepackage{enumitem}
\usepackage{mathrsfs}
\usepackage{lmodern}
\usepackage[T1]{fontenc}
\usepackage{dsfont}
\usepackage{cite}

\DeclareFontFamily{U}{mathc}{}
\DeclareFontShape{U}{mathc}{m}{it}%
{<->s*[1.03] mathc10}{}
\DeclareMathAlphabet{\mathlcal}{U}{mathc}{m}{it}

\numberwithin{equation}{section}

\providecommand{\hypersetup}[1]{}
\providecommand{\hyperref}[2][]{#2}

\providecommand{\pdfbookmark}[3][]{}

\hypersetup{plainpages=false}
\hypersetup{pdfpagemode=UseOutlines}
\hypersetup{bookmarksnumbered=true}
\hypersetup{bookmarksopen=true}
\hypersetup{pdfstartview=FitH}
\hypersetup{citecolor=[rgb]{0 .5 .5}}
\hypersetup{urlcolor=[rgb]{0.4 0 0}}
\hypersetup{linkcolor=[rgb]{1 0 1}}
\hypersetup{citebordercolor={.6 .9 .6}}
\hypersetup{urlbordercolor={1 0.75 0.8}}
\hypersetup{linkbordercolor={0 .6 .6}}
\hypersetup{pdfborder={0 0 .5}}

\newcommand{\namedref}[2]{\hyperref[#2]{#1~\ref*{#2}}}
\newcommand{\secref}[1]{\namedref{Section}{#1}}
\newcommand{\appref}[1]{\namedref{Appendix}{#1}}

\newcommand{\figref}[1]{\namedref{Figure}{#1}}

\usepackage{tikz}
\usepackage{pgfplots}
\usetikzlibrary{arrows,topaths,shapes.geometric,shapes.misc,patterns,positioning,shadows,decorations.markings,
decorations.pathreplacing,calc, trees, positioning, arrows, shapes, shapes.multipart, shadows, matrix, decorations.pathreplacing, decorations.pathmorphing}
\tikzset{->-/.style={decoration={
  markings,
  mark=at position #1 with {\arrow[scale=1.5]{latex}}},postaction={decorate}}}

\def\bra#1{\mathinner{\langle{#1}|}}
\def\ket#1{\mathinner{|{#1}\rangle}}

\newcommand{\ii}{{\rm i}}
\newcommand{\dd}{{\rm d}}

\def\be{\begin{equation}}
\def\ee{\end{equation}}
\def\bea{\begin{eqnarray}}
\def\eea{\end{eqnarray}}

\renewcommand\vec{\mathbf}

\newcommand{\ani}{\gamma}

\begin{document}

\begin{center}{\Large \textbf{Anisotropic Landau-Lifshitz Model in Discrete Space-Time}}\end{center}

\begin{center}
Žiga Krajnik\textsuperscript{1},
Enej Ilievski\textsuperscript{1}, 
Tomaž Prosen\textsuperscript{1}, and
Vincent Pasquier\textsuperscript{2}
\end{center}

\begin{center}
{\bf $^{1}$} Faculty of Mathematics and Physics, University of Ljubljana, Jadranska 19, 1000 Ljubljana, Slovenia\\
{\bf $^{2}$} Institut de Physique Th\'{e}orique, Universit\'{e} Paris Saclay, CEA, CNRS UMR 3681, 91191 Gif-sur-Yvette, France\\
{\small\ttfamily
~~~~\,\,{\href{mailto:ziga.krajnik@fmf.uni-lj.si}{ziga.krajnik@fmf.uni-lj.si}}
}
\end{center}

\begin{center}
\today
\end{center}


\section*{Abstract}{
We construct an integrable lattice model of classical interacting spins in discrete space-time, representing a discrete-time analogue of
the lattice Landau-Lifshitz ferromagnet with uniaxial anisotropy. As an application we use this explicit discrete symplectic integration scheme to compute the spin Drude weight and diffusion constant as functions of anisotropy and chemical potential. We demonstrate qualitatively different behavior in the
easy-axis and the easy-plane regimes in the non-magnetized sector. Upon approaching the isotropic point we also find an algebraic divergence of the diffusion constant, signaling a crossover to spin superdiffusion.
}

\vspace{10pt}
\noindent\rule{\textwidth}{1pt}
\tableofcontents\thispagestyle{fancy}
\noindent\rule{\textwidth}{1pt}
\vspace{10pt}

\section{Introduction}

The advent of powerful computational tools \cite{PhysRevLett.91.147902,PhysRevLett.93.207204} has tremendously advanced our understanding of
many-body physics in and out of equilibrium~\cite{RevModPhys.91.021001,NH_review,Bertini_review,ETH_review}.
Dealing with thermodynamic systems of strongly interacting degrees of freedom nonetheless still presents a dauting task in general.
In order to efficiently simulate many-body quantum dynamics on a computer one typically confronts the issue of growing complexity which arises when
distant regions of space become entangled. While \emph{classical} many-body dynamical systems on the other hand do not suffer this issue,
computing time-evolution of statistical ensembles often proves to be similarly challenging and require an immense amount of
computational resources.

Amidst recent experimental breakthroughs with cold-atom technologies~\cite{Bloch08,Bloch12}, there has been an upsurge of interest in the study of out-of-equilibrium phenomena. Nowadays, optical lattices not only enable manufacturing of tailor-made nanodevices for processing quantum information, but also offer a versatile platform to directly probe the relaxation process at finite energy density in the systems both near and far from equilibrium~\cite{Langen15,Schemmer19,Ketterle20,Malvania21}.

Concurrently, we have seen an increased theoretical interest in various facets of nonequilibrium physcis.
Integrable systems have been in the spotlight lately~\cite{DDKY17,Misguich17,Ljubotina17-diff,DYC18,SciPostPhys.2.2.014,PhysRevLett.120.045301,Bastianello_noise,CDDY19,BPK2019,KlumperDrude,BPP20}, largely due to their inherent non-ergodic features \cite{CCR11,VR_review,PhysRevLett.115.157201,EF_review}
and anomalous transport properties \cite{BCDF16,CDY16,Bulchandani17,LjubotinaNature17,Bulchandani18,LjubotinaPRL19,DupontMoore19,GV19,GVW19,NMKI20,Agrawal20,Bertini_review}
as recently covered in a compilation of review articles~\cite{GHDquenches_review,FF_review,diffusion_review,superdiffusion_review,RCA54_review,currents_review}.
The study of nonequilibrium properties in classical integrable dynamical systems of interacting particles or fields
has received comparatively less attention~\cite{Bastianello18,DoyonToda19,Gamayun19,SpohnToda20,SpohnJPA20,Kuniba_2020,Croydon_2020,kuniba2020generalized}.

One major point of worry concerning numerical simulations of \emph{integrable} dynamics is that integration schemes based on na\"{i}ve
(e.g. Trotter--Suzuki, or Runge--Kutta) discretizations invariably destroy integrability. On sufficiently short time scales this should not be
a major concern. By contrast, even weakly broken integrability is likely to induce spurious effects at late times and thus preclude reliable extraction of transport coefficients or dynamical exponents. In addition statistical field theories are commonly plagued by UV divergences (even when the local target space is a compact manifold). These drawbacks can both be obviated in a fully discrete setting.

Integrable many-body dynamical systems in discrete time \cite{VanicatTrotterization,krajnik2020integrable} and classical cellular
automata \cite{Klobas18,RCA54_review} have recently attracted much interest. An integrable space-time discretization of the isotropic classical Heisenberg ferromagnet has been recently obtained in Ref.~\cite{krajnik2020kardar},
and subsequently generalized to a large class of `matrix models' \cite{krajnik2020integrable} which are globally invariant under the
action of simple Lie groups. In this work, we construct an \emph{anisotropic} deformation of the classical $SU(2)$ ferromagnet obtained in Ref.~\cite{krajnik2020kardar}, representing a `brick-wall type' circuit composed from elementary two-body symplectic maps. Our model
can be thus regarded as an `integrable Trotterization' \cite{GP17,ljubotina2019ballistic,krajnik2020kardar,krajnik2020integrable}
of the anisotropic lattice Landau-Lifshitz field theory~\cite{Takhtajan77,Sklyanin79,Faddeev1987} (the classical counterpart of the celebrated quantum XXZ spin chain). A fully discrete integrable Landau--Lifshitz equation has, to the best of our knowledge, not been constructed yet;
it has only appeared previously in a paper by Hirota \cite{Hirota82} in a bilinear form in terms of the discrete {\em tau} function.
The outcome of our construction is an explicit finite-step integrable integration scheme that can be efficiently employed to computationally study dynamical properties of the model. As an application, we compute the Drude weight and diffusion constant as functions of spin chemical potential (i.e. magnetization) and anisotropy.

\paragraph*{Outline.}
This paper is structured as follows. We begin in \secref{sec:discrete_model} by describing the formal setting. First, we introduce
a discrete zero-curvature property of an auxiliary linear transport problem on a (tilted, light-cone) space-time lattice and interpret it in terms of a local dynamical map acting on a pair of classical spins. We proceed by introducing a dynamical system on a discrete space-time lattice in the form of a `symplectic circuit'.
Next, in \secref{sec:Lax} we define the local phase space equipped with a Poisson structure, and introduce the Lax matrix of the
lattice Landau--Lifshitz model. Moving on, in \secref{sec:factorization} we outline how to explicitly solve
the zero-curvature relation by exploiting the underlying algebraic relations, yielding the local time-propagator of the model.
\secref{sec:transport} is devoted to a physics application, where we carry out a detailed numerical study of magnetization transport
in grand-canonical Gibbs states for the easy-axis and easy-plane regimes of the model.
In \secref{sec:conclusion} we conclude with a brief summary of the main results.

\section{Anisotropic symplectic spin model in discrete space-time}
\label{sec:discrete_model}

Before delving into the specifics of the model, we first introduce the general setting.
We shall mostly follow the lines and use the notation from previous
works on related discrete-time models with isotropic interactions \cite{krajnik2020kardar,krajnik2020integrable}.

\paragraph{Discrete space-time.}
The physical space-time lattice is a two-dimensional square lattice with nodes $(\ell,t)\in \mathbb{Z}^{2}$, with $\ell$ and $t$
referring to the spatial and temporal indices, respectively. Throughout the paper we adopt the convention that time flows upwards while the
spatial axis is oriented towards the right. Moreover, we adopt periodic boundary conditions in space by identifying
$\ell\equiv \ell + L$ and additionally assume, for definiteness, the system length $L$ to be even.

Each site of the space-time lattice is attached a local physical degree of freedom
${\bf S}^{t}_{\ell}$, which we consider to be a classical spin $\vec{S} = (S^1, S^2, S^3)$ of unit length,
$\vec{S} \cdot \vec{S} = S^{+}S^{-} + (S^{3})^{2} = 1$ (with $S^{\pm}\equiv S^1 \pm \ii S^2$) obeying the canonical
ultralocal Lie--Poisson brackets
\begin{equation}
\{S^{a}_\ell, S^{b}_{\ell'}\} = \delta_{\ell, \ell'}\sum_c \epsilon_{abc} S^{c}_{\ell},
\label{canonical_Poisson_ultra_local}
\end{equation}
where we have used the Levi--Civita symbol $\epsilon_{abc}$.

We furthermore introduce a square light-cone lattice by tilting the space-time lattice by $45^{\circ}$ degrees, and
assign to its vertices (nodes) $(n,m)\in \mathbb{Z}^{2}$ auxiliary variables $\phi_{n,m}$. Physical spins accordingly sit
at the middle of the edges on the light-cone lattice.

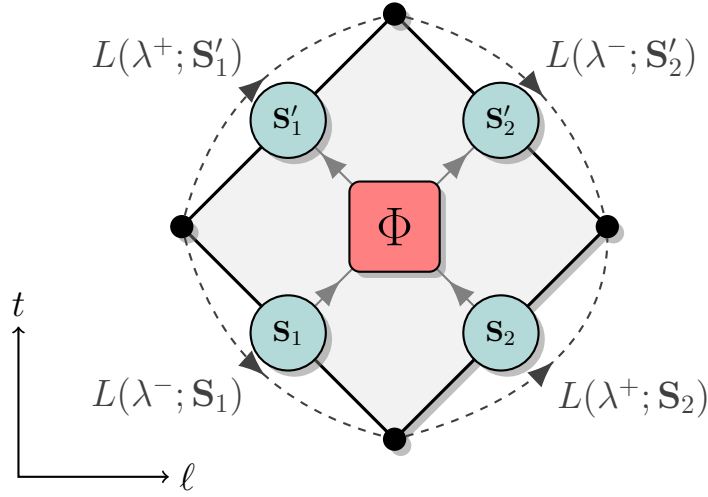
\begin{figure}[h]
\centering
\begin{tikzpicture}[scale=1,thick]

\draw [<->,thick] (-5,1.5) node (yaxis) [above] {\Large{$t$}}
        |- (-3,-0.5) node (xaxis) [right] {\Large{$\ell$}};

\begin{scope}[rotate=45]

\draw[fill=black!5,very thick, drop shadow] (0,0) rectangle (4,4);


\draw[gray,decoration={markings,mark=at position 0.5 with
    {\arrow[scale=2,>=stealth']{latex}}},postaction={decorate}] (0,2) -- (2,2);
\draw[gray,decoration={markings,mark=at position 0.7 with
    {\arrow[scale=2,>=stealth']{latex}}},postaction={decorate}] (2,2) -- (4,2);
\draw[gray,decoration={markings,mark=at position 0.5 with
    {\arrow[scale=2,>=stealth']{latex}}},postaction={decorate}] (2,0) -- (2,2);
\draw[gray,decoration={markings,mark=at position 0.7 with
    {\arrow[scale=2,>=stealth']{latex}}},postaction={decorate}] (2,2) -- (2,4);

\node at (0,0) [fill=black,inner sep=0pt,minimum size=8pt,draw,circle, drop shadow] (b) {};
\node at (4,0) [fill=black,inner sep=0pt,minimum size=8pt,draw,circle, drop shadow] (r) {};
\node at (0,4) [fill=black,inner sep=0pt,minimum size=8pt,draw,circle, drop shadow] (l) {};
\node at (4,4) [fill=black,inner sep=0pt,minimum size=8pt,draw,circle, drop shadow] (t) {};
\node at (0,2) [fill=teal!30,minimum size= 1cm,draw,circle,drop shadow] (M1) {$\vec{S}_{1}$};
\node at (2,0) [fill=teal!30,minimum size= 1cm,draw,circle,drop shadow] (M2) {$\vec{S}_{2}$};
\node at (2,4) [fill=teal!30,minimum size= 1cm,draw,circle,drop shadow] (M1p) {$\vec{S}^{\prime}_{1}$};
\node at (4,2) [fill=teal!30,minimum size= 1cm,draw,circle,drop shadow] (M2p) {$\vec{S}^{\prime}_{2}$};

\node at (2,2) [fill=red!50,minimum size=1.2cm,draw,rounded corners, drop shadow] (P1i) {\huge{$\Phi$}};

\end{scope}

\draw[-,darkgray,dashed,decoration={markings,mark=at position 0.55 with
    {\arrow[scale=2,>=stealth']{latex}}},postaction={decorate}] (l) to [out=-75,in=165] node [below left] {\Large{$L(\lambda^-;\vec{S}_{1})$}} (b);

\draw[-,darkgray,dashed,decoration={markings,mark=at position 0.55 with
    {\arrow[scale=2,>=stealth']{latex}}},postaction={decorate}] (b) to [out=15,in=-90] node [below right] {\Large{$L(\lambda^+;\vec{S}_{2})$}} (r);

\draw[-,darkgray,dashed,decoration={markings,mark=at position 0.55 with
    {\arrow[scale=2,>=stealth']{latex}}},postaction={decorate}] (l) to [out=75,in=-165] node [above left] {\Large{$L(\lambda^+;\vec{S}^{\prime}_{1})$}} (t);

\draw[-,darkgray,dashed,decoration={markings,mark=at position 0.55 with
    {\arrow[scale=2,>=stealth']{latex}}},postaction={decorate}] (t) to [out=-15,in=105] node [above right] {\Large{$L(\lambda^-;\vec{S}^{\prime}_{2})$}} (r);

\end{tikzpicture}
\caption{Elementary plaquette of the discrete light-cone lattice: classical spins ${\bf S}\in S^{2}$ (circles at the middle of the edges)
are situated on the vertices of the discrete space-time lattice. Primed variables $\vec{S}^{\prime}_{1,2}$ pertain
to spins $\vec{S}_{1,2}$ that are time-shifted forward by one unit of time by applying the propagator $\Phi$ (red square).}
\label{fig:plaquette}
\end{figure}

\paragraph{Discrete zero-curvature condition.}
Integrabiliy of a dynamical systems in discrete time is a direct manifestation of a discrete zero-curvature property of an associated
linear transport problem~\cite{Lax1968,AKNS1974} for auxiliary variables $\phi$, in analogy to completely integrable
Hamiltonian systems (see e.g. Refs.~\cite{Faddeev1987,AblowitzSegur_book,Babelon_book}). To realize parallel transport on the auxiliary
light-cone lattice we introduce the lattice shift operators \cite{Hietarinta_book} along the light-cone directions
(i.e. characteristics $\ell \pm t = {\rm const}$)
\begin{equation}
\phi_{n+1,m} = L^{(+)}_{n,m}(\lambda)\phi_{n,m},\qquad \phi_{n,m+1} = L^{(-)}_{n,m}(\lambda)\phi_{n,m}.
\label{linear_problem}
\end{equation}
Here the local `matrix propagators' $L^{(\pm)}$ represent certain matrix-valued functionals of physical variables which, in addition,
depend analytically on a free complex parameter $\lambda$.  In order to ensure consistency of such parallel transport,
there should be no ambiguity in the order of propagation when passing from $\phi_{n,m}$ to $\phi_{n+1,m+1}$.
In this case, the shift matrices $L^{(\pm)}$ constitute the \emph{Lax pair} and $\lambda$ is called the spectral parameter.
More generally, one can also allow the spectral parameter to have additional dependence on local `inhomogeneities' along an initial sawtooth.
We shall make a special `homogeneous choice' by requesting $L^\pm\equiv L(\lambda^{\pm}; \vec{S})$, i.e. using a pair of spectral parameters
$\lambda^{\pm}$ depending only on the light-cone direction (but not on position $\ell$ or time $t$).

Commutativity of the light-cone shifts is neatly encapsulated by the following \emph{discrete zero-curvature property} \cite{Moser_1991,Veselov_2003} around
an elementary square plaquette of the light-cone lattice,\footnote{A magnetic field could be incorporated in the construction by adding constant `twist' matrices to the zero-curvature condition. See \cite{krajnik2020integrable} for a similar construction.}
\begin{equation}
L(\lambda^+;\vec{S}_{2})L(\lambda^-;\vec{S}_{1})
= L(\lambda^-;\vec{S}^{\prime}_{2})L(\lambda^+;\vec{S}^{\prime}_{1}).
\label{eqn:ZC_abstract}
\end{equation}
where we have assumed, for notational convenience, that the `input' variables $\vec{S}_{1,2}$ sit at lattice sites $1$ and $2$.
Notice that each of the light-cone Lax matrices $L(\lambda^{\pm})$ depend on the local `edge variable' $\vec{S}$ on the corresponding edge,
whereas primed variables $\vec{S}^{\prime}$ are being used as a shorthand notation for the propagated spins, i.e. original variables
$\vec{S}$ shifted by one unit in the time direction (as depicted in \figref{fig:plaquette}). 
The discrete curvature is fulfiled \emph{if and only if} the time-updated `output' variables $\vec{S}^{\prime}_{1,2}$ are appropriately
linked to `input' variables $\vec{S}_{1,2}$. This requirement allows us to interpret the flatness condition as an (implicit)
specification of a local time-propagator, i.e. a local two-body map $\Phi: (\vec{S}_1,\vec{S}_2)\mapsto(\vec{S}^{\prime}_1,\vec{S}^{\prime}_2)$
for any spatially adjacent pair of sites.

\medskip
The zero-curvature condition implies the integrability of the model through a property called \emph{isospectrality}.
As explained in \appref{sec:iso}, isospectrality gives rise to an infinite set of local conserved quantities in involution.

\paragraph{Many-body propagator.}

Let us for the time being leave the two-body propagator $\Phi$ defined implicitly via the zero-curvature property. In the following,
we first outline how to use $\Phi$ as the elementary building block of a many-body dynamical map in the form of a `brick-wall circuit'.
To this end, let $\mathcal{M}_{1} \cong S^{2}$ denote a local phase space (a $2$-sphere) of a canonical spin.
The full phase space $\mathcal{M}_{L}$ of an $L$-site chain is thus given simply by the $L$-fold Cartesian product,
$\mathcal{M}_{L}\equiv \mathcal{M}^{\times L}_{1}$, and accordingly we introduce a many-body dynamical map
$\Phi^{\rm full}:\mathcal{M}_{L}\mapsto \mathcal{M}_{L}$.

\medskip

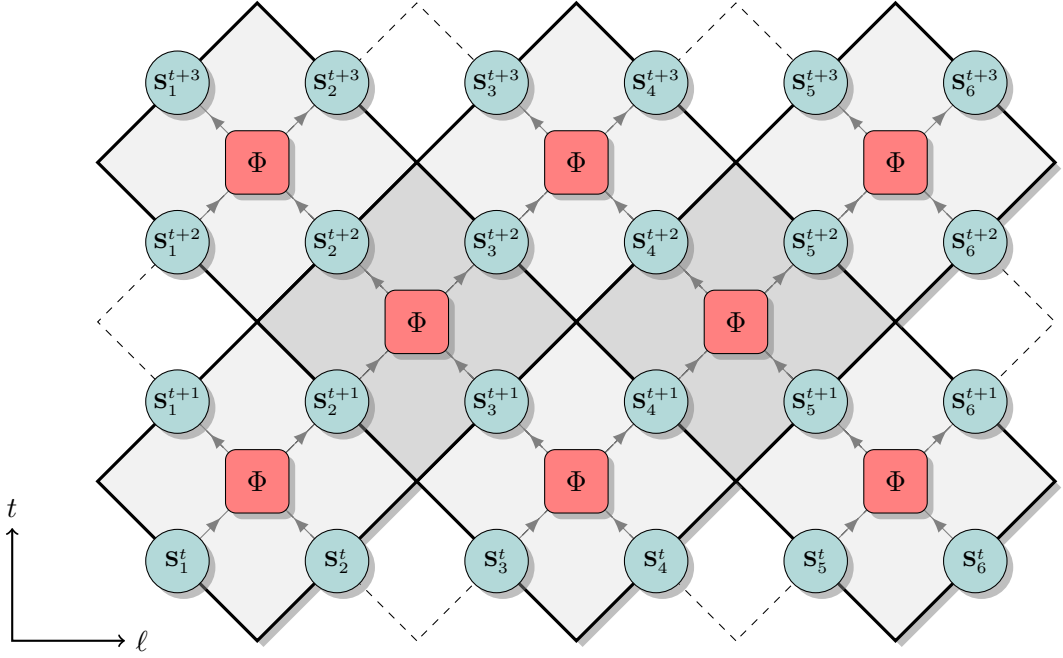
\begin{figure}[t]
\centering
\begin{tikzpicture}[scale=1.5]
\draw [<->,thick] (-5,1) node (yaxis) [above] {$t$}
        |- (-4,0) node (xaxis) [right] {$\ell$};

\begin{scope}[rotate=45]
\draw[fill=black!5,very thick, drop shadow] (-2,2) rectangle (0,4);
\draw[fill=black!5,very thick, drop shadow] (2,-2) rectangle (4,0);
\draw[fill=black!5,very thick, drop shadow] (0,0) rectangle (2,2);
\draw[fill=black!5,very thick, drop shadow] (2,2) rectangle (4,4);
\draw[fill=black!5,very thick, drop shadow] (0,4) rectangle (2,6);
\draw[fill=black!5,very thick, drop shadow] (4,0) rectangle (6,2);

\draw[fill=black!15,very thick] (2,0) rectangle (4,2);
\draw[fill=black!15,very thick] (0,2) rectangle (2,4);

\draw[dashed] (-1,1) rectangle (1,3);
\draw[dashed] (1,-1) rectangle (3,1);
\draw[dashed] (1,1) rectangle (3,3);
\draw[dashed] (3,-1) rectangle (5,1);
\draw[dashed] (-1,3) rectangle (1,5);
\draw[dashed] (3,1) rectangle (5,3);
\draw[dashed] (1,3) rectangle (3,5);

\foreach \i in {1,...,3} {
	\draw[gray,decoration={markings,mark=at position 0.65 with
    		{\arrow[scale=1.5,>=stealth']{latex}}},postaction={decorate}] (-3+2*\i,5-2*\i) -- (-3+2*\i+1,5-2*\i);
    	\draw[gray,decoration={markings,mark=at position 0.55 with
    		{\arrow[scale=1.5,>=stealth']{latex}}},postaction={decorate}] (-3+2*\i-1,5-2*\i) -- (-3+2*\i,5-2*\i);
	\draw[gray,decoration={markings,mark=at position 0.55 with
    		{\arrow[scale=1.5,>=stealth']{latex}}},postaction={decorate}] (-3+2*\i,5-2*\i-1) -- (-3+2*\i,5-2*\i);
	\draw[gray,decoration={markings,mark=at position 0.6 with
    		{\arrow[scale=1.5,>=stealth']{latex}}},postaction={decorate}] (-3+2*\i,5-2*\i) -- (-3+2*\i,5-2*\i+1);

	\node at (-3+2*\i,5-2*\i) [fill=red!50,inner sep=0pt,minimum size=24pt,draw,rounded corners, drop shadow] (P1i) {$\Phi$};

	\draw[gray,decoration={markings,mark=at position 0.65 with
    		{\arrow[scale=1.5,>=stealth']{latex}}},postaction={decorate}] (-1+2*\i,7-2*\i) -- (-1+2*\i+1,7-2*\i);
    	\draw[gray,decoration={markings,mark=at position 0.55 with
    		{\arrow[scale=1.5,>=stealth']{latex}}},postaction={decorate}] (-1+2*\i-1,7-2*\i) -- (-1+2*\i,7-2*\i);
	\draw[gray,decoration={markings,mark=at position 0.55 with
    		{\arrow[scale=1.5,>=stealth']{latex}}},postaction={decorate}] (-1+2*\i,7-2*\i-1) -- (-1+2*\i,7-2*\i);
	\draw[gray,decoration={markings,mark=at position 0.6 with
    		{\arrow[scale=1.5,>=stealth']{latex}}},postaction={decorate}] (-1+2*\i,7-2*\i) -- (-1+2*\i,7-2*\i+1);

	\node at (-1+2*\i,7-2*\i) [fill=red!50,inner sep=0pt,minimum size=24pt,draw,rounded corners, drop shadow] (P1i) {$\Phi$};
}

\foreach \i in {1,...,2} {

	\draw[gray,decoration={markings,mark=at position 0.65 with
    		{\arrow[scale=1.5,>=stealth']{latex}}},postaction={decorate}] (-1+2*\i,5-2*\i) -- (-1+2*\i+1,5-2*\i);
    	\draw[gray,decoration={markings,mark=at position 0.55 with
    		{\arrow[scale=1.5,>=stealth']{latex}}},postaction={decorate}] (-1+2*\i-1,5-2*\i) -- (-1+2*\i,5-2*\i);
	\draw[gray,decoration={markings,mark=at position 0.55 with
    		{\arrow[scale=1.5,>=stealth']{latex}}},postaction={decorate}] (-1+2*\i,5-2*\i-1) -- (-1+2*\i,5-2*\i);
	\draw[gray,decoration={markings,mark=at position 0.6 with
    		{\arrow[scale=1.5,>=stealth']{latex}}},postaction={decorate}] (-1+2*\i,5-2*\i) -- (-1+2*\i,5-2*\i+1);

	\node at (-1+2*\i,5-2*\i) [fill=red!50,inner sep=0pt,minimum size=24pt,draw,rounded corners, drop shadow] (P1i) {$\Phi$};
}

\foreach \i in {1,...,6} {

	\node at (-3+\i,4-\i) [fill=teal!30,inner sep=0pt,minimum size=24pt,draw,circle,drop shadow] (M1i) {\scriptsize{$\vec{S}^{t}_{\i}$}};
	\node at (-3+\i+1,4-\i+1) [fill=teal!30,inner sep=0pt,minimum size=24pt,draw,circle,drop shadow] (M2i) {\scriptsize{$\vec{S}^{t+1}_{\i}$}};
	\node at (-3+\i+2,4-\i+2) [fill=teal!30,inner sep=0pt,minimum size=24pt,draw,circle,drop shadow] (M3i) {\scriptsize{$\vec{S}^{t+2}_{\i}$}};
	\node at (-3+\i+3,4-\i+3) [fill=teal!30,inner sep=0pt,minimum size=24pt,draw,circle,drop shadow] (M4i) {\scriptsize{$\vec{S}^{t+3}_{\i}$}};
}

\end{scope}

\end{tikzpicture}
\caption{Fabric of discrete space-time: the square space-time lattice comprising classical spin degrees of freedom
$\vec{S}^{t}_{\ell}$ (circles) superimposed on a square light-cone lattice represented by a tilted checkerboard.
In the middle of each plaquette, there is a two-body symplectic map $\Phi$ (red square) which maps the pairs of physical spin variables
forward in time.}
\label{fig:circuit}
\end{figure}

By virtue of the light-cone geometry, the full time-dynamics consists of an alterating sequence of `odd' and `even' local propagators,
\begin{equation}
(\vec{S}^{2t+2}_{2\ell-1},\vec{S}^{2t+2}_{2\ell}) = \Phi(\vec{S}^{2t+1}_{2\ell-1},\vec{S}^{2t+1}_{2\ell}),\qquad
(\vec{S}^{2t+1}_{2\ell},\vec{S}^{2t+1}_{2\ell+1}) = \Phi(\vec{S}^{2t}_{2\ell},\vec{S}^{2t}_{2\ell+1}),
\label{eqn:st}
\end{equation}
respectively. This results in a staggered circuit as depicted in \figref{fig:circuit}. By making use of the embedding prescription,
\begin{equation}
\Phi^{(j)} = I^{\times (j-1)}\,\times \, \Phi\, \times\, I^{\times (L-j-1)},
\end{equation}
for $j=1,\ldots,L-1$, where $I:\mathcal{M}_{1}\mapsto \mathcal{M}_{1}$ stands for a local (single-site) unit map $I(\vec{S})\equiv \vec{S}$,
($\Phi^{(L)}$ correspondingly affects the $L$th and the first spin), the full propagator $\Phi^{\rm full}$ for two units of time $t\mapsto t+2$ decomposes as
\begin{equation}
\Phi^{{\rm full}} = \Phi^{\rm even}\circ \Phi^{\rm odd},
\label{full_propagator}
\end{equation}
with odd and even propagators further factorizing into commuting two-body maps,
\begin{equation}
\Phi^{\rm odd} = \prod_{\ell=1}^{L/2} \Phi^{(2\ell)}, \qquad
\Phi^{\rm even}= \prod_{\ell=1}^{L/2} \Phi^{(2\ell-1)}.
\end{equation}

\medskip

So far our construction has been entirely formal. In what follows, we derive the map $\Phi^{\rm full}$ that corresponds to an integrable
space-time discretization of the anisotropic lattice Landau--Lifshitz model, as introduced by Sklyanin \cite{Sklyanin79,Sklyanin83,Faddeev1987}.
In this view, $\Phi^{\rm full}$ represents its `integrable Trotterrization'.

\subsection{Sklyanin Lax matrix}
\label{sec:Lax}

In this section, we construct an integrable time-discrete analogue of the anisotropic
lattice Landau--Lifshitz model. To facilitate the computations, instead of canonical spins ${\bf S}$, it proves more convenient to employ
an axially `deformed spin' with components $\{\mathcal{K},\mathcal{S}^{+},\mathcal{S}^{-}\}$, enclosing a Poisson algebra
\begin{equation}
\mathcal{A}_{q}:\qquad \{\mathcal{K}, \mathcal{S}^{\pm}\} = \mp \ii \varrho \, \mathcal{S}^{\pm} \mathcal{K}, \qquad
\{\mathcal{S}^{+}, \mathcal{S}^{-}\} = -\frac{ \ii \varrho}{2}\left(\mathcal{K}^{2} - \mathcal{K}^{-2}\right).
\label{Sklyanin_bracket}
\end{equation}
This Poisson algebra admits a quadric (Casimir) function
\begin{equation}
\mathcal{C}_{0} \equiv \mathcal{S}^{+} \mathcal{S}^{-} + \left(\frac{\mathcal{K}-\mathcal{K}^{-1}}{2}\right)^{2},
\label{anisotropic_sphere}
\end{equation}
satisfying $\{\mathcal{C}_{0},\mathcal{S}^{a}\}=0$.  In terms of `Sklyanin spins'~\footnote{Classical canonical spins can be
recovered in the $\varrho \to 0$ limit by simultaneously rescaling $\mathcal{S}^{a}\to \mathcal{S}^{a}/\varrho$ for $a\in \{1,2,3\}$.},
with components $\{\mathcal{S}^{a}\}_{a=0}^{3}$, with
\begin{equation}
\mathcal{S}^{0}\equiv \frac{1}{2}(\mathcal{K}+\mathcal{K}^{-1}),\quad
\mathcal{S}^{1}\equiv \frac{1}{2}(\mathcal{S}^{+} + \mathcal{S}^{-}),\quad 
\mathcal{S}^{2}\equiv \frac{1}{2\ii}(\mathcal{S}^{+} - \mathcal{S}^{-}),\quad
\mathcal{S}^{3}\equiv \frac{1}{2}(\mathcal{K}-\mathcal{K}^{-1}),
\end{equation}
the Hamiltonian of the lattice Landau--Lifshitz model takes a compact form
\begin{equation}
H_{\rm LLL}^{\varrho} \simeq \sum_{\ell=1}^{L} \log \left[ \sum_{a=0}^{3}g_{a}\mathcal{S}^{a}_{\ell}\mathcal{S}^{a}_{\ell+1}\right],
\label{LLL_Hamiltonian}
\end{equation}
with couplings $g_{0}=\sinh^{2}{(\varrho)}$, $g_{1}=g_{2}=1$ and $g_{3}=\cosh^{2}{(\varrho)}$.

We note that $\mathcal{A}_{q}$  is just the trigonometric limit of `elliptic spins' that constitute
the so-called quadratic \emph{Poisson--Sklyanin algebra}, cf. \appref{app:semiclassical} for further information.
The algebra $\mathcal{A}_{q}$ becomes non-degenerate on symplectic leaves on which $\mathcal{C}_{0}$ takes a constant value $c_{0}$.
We consider two \emph{physically distinct} regimes:
\begin{enumerate}[label=(\roman*)]
\item the \emph{easy-axis} regime, with a real anisotropy (or interaction) parameter $\varrho \in \mathbb{R}_{+}$,
\item the \emph{easy-plane} regime, with an interaction parameter $\ani$ having a \emph{compact support}~\footnote{This restriction has been
overlooked in the previous works \cite{PZ13,DasKPZ} on the lattice Landau--Lifshitz model in continuous time, which specialize to the value $\ani=1$.}
$\ani \in [-\pi/2,\pi/2]$. The associated Hamiltonian is obtained from Eq.~\eqref{LLL_Hamiltonian} by analytically continuing $\varrho$ from
the real axis onto the imaginary axis, $\varrho \to \ii \ani$.
\end{enumerate}

Our construction, which we outline in turn, applies to both of these regimes. Before that, we first briefly discuss the
symplectic structure of the local space space. To obtain the easy-axis lattice Landau--Lifshitz model,
the Casimir function $\mathcal{C}_{0}$ has to be fixed to
\begin{equation}
c_{0} = \sinh^{2}{(\varrho)}.
\label{Casimir_value}
\end{equation}
In this way, we select a two-dimensional non-degenerate Poisson submanifold $\mathcal{M}_{\rm ea}$ inside $\mathbb{R}^{3}$
which is diffeomorphic to a $2$-sphere. In the easy-plane regime, obtained under analytic continuation $\varrho \to \ii \gamma$,
the submanifold $\mathcal{M}_{\rm ep}$ is also topologically equivalent to a $2$-sphere provided $\ani \in [-\pi/2,\pi/2]$.
The upshot here is that there exist a smooth bijective mapping to the symplectic sphere $S^{2}$, implying that
variables $\mathcal{K}$ and $\mathcal{S}^{\pm}$ can be parameterized in terms of classical \emph{canonical} spins ${\bf S}$.
A local mapping from $S^{2}$ to $\mathcal{M}_{\rm ea}$ is given by \cite{Faddeev1987}
\begin{equation}
\mathcal{K} = e^{\varrho \, S^{3}}, \qquad
\mathcal{S}^{\pm} =  F_{\varrho}(S^{3})S^{\pm},
\label{Sklyanin_vars}
\end{equation}
where
\begin{equation}
F_\varrho(s) \equiv \sqrt{\frac{\sinh^{2}(\varrho) - \sinh^2(\varrho s)}{1-s^2}},
\end{equation} 
the form of which follows from the Casimir function, see Eqs.~\eqref{anisotropic_sphere},\eqref{Casimir_value}.
Two imporatnt remarks are in order at this stage: (I) in the \emph{easy-plane regime} (with $\varrho \to \ii \ani$
and $F_{\ii\ani}(s)$), Eq.~\eqref{Sklyanin_vars} provides a \emph{bijective} map only inside a compact interval of anisotropies
$|\ani|\leq \pi/2$; (II) there exist other Poisson submanifolds (classified in \cite{Faddeev1987,Sklyanin89XYZ}) where the classical Sklyanin algebra becomes non-degenerate. Those are not expressible in terms of canonical spins and thus we do not consider them here.

\medskip

The discrete zero-curvature condition \eqref{eqn:ZC_abstract} may be formally viewed as a matrix refactorization problem.
The task ahead of us is to find its explicit solution, i.e. expressing the output (primed) variables in terms of the input spin variables.
We achieve this by first casting the Sklyanin Lax matrix in the form
\begin{equation}
L(z) =\frac{2}{z - z^{-1}}
\begin{bmatrix}
\tfrac{1}{2} \left( z \, \mathcal K - z^{-1} \mathcal K^{-1} \right)   &  \mathcal S^-\\
\mathcal S^+ &  -\tfrac{1}{2} \left(z^{-1}\mathcal K - z \, \mathcal K^{-1} \right)
\end{bmatrix},
\label{Lax1}
\end{equation}
where we have for convenience introduced a multiplicative spectral parameter $z \in \mathbb{C}$,
\begin{equation}
z(\lambda) = e^{\ii \varrho \lambda}.
\label{z_lam_param}
\end{equation}
In \appref{app:semiclassical} we detail out how the above Sklyanin Lax matrix arises from the limit of the quantum Lax operator associated to the quantum Yang--Baxter algebra.

The task of refactorization is unfortunately not straightforward. To begin with, Eq.~\eqref{eqn:ZC_abstract} merely provides an implicit rule for
the time-propagated input variables, and there is no obvious way of recasting it as an explicit map $\Phi$.~\footnote{This is to some
extent true even in the simplest isotropic case, where nonetheless it is possible to make use of vector identities to solve the zero-curvature condition explicitly.}
This difficulty can be elegantly overcome by exploting certain factorization properties of the Lax matrix into more fundamental
constituents.

\subsection{Factorization}
\label{sec:factorization}

In this section, we outline how the Lax matrix can be further decomposed into more elementary constituents.
Such factorization properties are indeed deepely rooted in the algebraic properties of certain types of Hopf algebras that are
associated to simple Lie algebras or deformations thereof. For instance, it is well known that the universal
\emph{quantum} $\mathcal{R}$-matrix admits a triangular (Borel) decomposition \cite{FRT88,Drinfeld88,KT96,KT14}.
The resulting `partonic' Lax operators can be used to construct the Baxter $Q$-operators \cite{Baxter_book}.
There exist various equivalent constructions in the literature,
see e.g. Refs.~ \cite{DerkachovI,DerkachovII,Derkachov06} and Refs.~\cite{Bazhanov10,Frassek11,Babelon_2018}. We in turn demonstrate how
the entire construction quite naturally elevates to the classical level by considering its semiclassical limit (outlined
in \appref{app:semiclassical}). Below we collect the relevant formulae without delving too much into formal considerations.

\paragraph{Weyl variables.}
To facilitate the factorization of the Lax matrix, we parametrize Sklyanin variables in terms of a classical Weyl--Poisson algebra.
The latter is spanned by a pair of generators, $x$, $y$, satisfying the multiplicative canonical Poisson bracket
\begin{equation}
\{x, y\} = \ii \varrho\, x\,y,
\label{Weyl_bracket}
\end{equation}
via the prescription
\begin{equation}
\mathcal{K} = \nu \, y, \qquad
\mathcal{S}^{+} = \frac{x}{2} \left(\nu^{2} y - \nu^{-2}y^{-1} \right), \qquad
\mathcal{S}^{-} = - \frac{1}{2x} \left(y - y^{-1} \right),
\label{Weyl}
\end{equation}
where we have simultaneously defined, for compactness of notation, $\nu \equiv e^{\varrho}$.
We accordingly make the identification
\begin{equation}
y = \nu^{-1}\mathcal{K} = e^{\varrho\,(S^{3}-1)}.
\label{y_sz}
\end{equation}
This implies that $y$ is real and positive in the easy-axis case, whereas in the easy-plane regime it becomes unimodular.
In Weyl variables, however, the reality condition $S^{+} = \overline{S^{-}}$ is not manifestly satisfied.
To enforce it, the variable $x$ has to be restricted to a `physical' submanifold by demanding the squared modulus to obey
\begin{equation}
|x|^{2} = -
[y]_{\nu^{2}}, 
\qquad
[y]_{\nu^{2}}\equiv \frac{y-y^{-1}}{\nu^{2}y-\nu^{-2}y^{-1}}. \label{reality_condition}
\end{equation}

\paragraph*{Factorization.}
By introducing\footnote{An overall multiplicative scalar factor bears no consequence on the dynamics as it cancels out from the discrete zero-curvature condition \eqref{eqn:ZC_abstract}.}
\begin{equation}
L_{x,y} \equiv \frac{1}{2}\left(z-z^{-1}\right)L(z),
\end{equation}
along with a pair of spectral parameters,
\begin{equation}
u = \nu \, z, \qquad v = \nu \, z^{-1},
\end{equation}
satisfying $uv = \nu^2$, $u/v = z^2$, the Lax operator from Eq.~\eqref{Lax1} can be factorized as
\begin{equation}
L_{x,y}(V,U) \equiv \frac{1}{2} X^{-1}VYU^T X =  \frac{1}{2}  \begin{bmatrix}
uy - u^{-1}y^{-1} & -(y-y^{-1})x^{-1}\\
(uvy - u^{-1}v^{-1}y^{-1})x & - vy + v^{-1}y^{-1}
\end{bmatrix} ,
\end{equation} 
in terms of diagonal `coordinate' matrices $X$ and $Y$
\begin{equation}
X = \begin{bmatrix}
x & 0 \\
0 & 1
\end{bmatrix}, \qquad
Y = \begin{bmatrix}
y & 0 \\
0 & y^{-1}
\end{bmatrix},
\end{equation}
and the `spectral' matrices $U$ and $V$
\begin{equation}
U = \begin{bmatrix}
u & - u^{-1}\\
-1 & 1
\end{bmatrix}, \qquad
V = \begin{bmatrix}
1 & 1\\
v & v^{-1}
\end{bmatrix}.
\end{equation}

\paragraph{Elementary exchange relations.}
We next describe various elementary exchange procedures involving $L_{x,y}(U,V)$. Firstly, interchanging the spectral matrices
\begin{equation}
L_{x, y} (V, U) = L_{x', y'}(U, V),
\end{equation}
specifies a map $(x, y) \mapsto (x', y')$. The canonical Weyl relations are preserved~\footnote{This can be inferred directly using
that $x' = x\,f(y)$ and $y' = y$ preserve the Poisson bracket \eqref{Weyl_bracket}.} by additionally assuming $y'=y$, in which case
\begin{equation}
X^{-1}VYU^T X = X'^{-1}UYV^T X'.
\label{eq1}
\end{equation}
Secondly, there exist quadratic exchange relations of the type
\begin{align}
L_{x_1, y_1}(U_1, V_1) L_{x_2, y_2}(V_2, U_2) &= L_{x'_1, y'_1}(U_1, V_2) L_{x'_2, y'_2}(V_1, U_2),\\
L_{x_1, y_1}(V_1, U_1) L_{x_2, y_2}(U_2, V_2) &= L_{x'_1, y'_1}(V_1, U_2) L_{x'_2, y'_2}(U_1, V_2), 
\end{align}
which are canonical provided that $x_1 = x_1'$, $x_2' = x_2$, implying
\begin{align}
Y_1 V_1^T X_1 X_2^{-1} V_2 Y_2 &= Y_1' V_2^T X_1 X_2^{-1} V_1 Y_2', \label{eq2}\\
Y_1 U_1^T X_1 X_2^{-1} U_2 Y_2 &= Y_1' U_2^T X_1 X_2^{-1} U_1 Y_2'. \label{eq3}
\end{align}
From Eq.~\eqref{eq1} we can deduce the following transformation of $(x, y) \mapsto (x', y'=y)$:
\begin{equation}
x' = -x \frac{uvy - (uvy)^{-1}}{y - y^{-1}}.
\end{equation}
Equations \eqref{eq2} and \eqref{eq3} are equivalent under interchaging $u \leftrightarrow v$, so it suffices to solve only one of them. 
By direct calculation, exchanging $V_1$ and  $V_2$ according to Eq.\eqref{eq2} yields
\begin{equation}
(y_1')^2 = y_1^2 \frac{v_1}{v_2} \frac{v_1 x_2 + v_2 x_1}{v_1 x_1 + v_2 x_2}, \qquad
(y_2')^2 = y_2^2 \frac{v_2}{v_1} \frac{v_1 x_1 + v_2 x_2}{v_1 x_2 + v_2 x_1}, \qquad
x_\ell' = x_\ell,
\label{eq4}
\end{equation}
$\ell\in\{1,2\},$
while analogously the exchange of $U_1$ and $U_2$ according to Eq.~\eqref{eq3} yields the same expression with $v_\ell$ replaced by $u_\ell$:
\begin{equation}
(y_1')^2 = y_1^2 \frac{u_1}{u_2} \frac{u_1 x_1 + u_2 x_2}{u_2 x_1 + u_1 x_2}, \quad (y_2')^2 = y_2^2 \frac{u_2}{u_1} \frac{u_2 x_1 + u_1 x_2}{u_1 x_1 + u_2 x_2}, \quad x_\ell' = x_\ell.
\label{eq5}
\end{equation}
By the same reasoning as previously (only swapping the roles of $x_\ell$ and $y_\ell$), transformations \eqref{eq4} and \eqref{eq5} are both canonical.

\paragraph{Composite exchange relations.}
By using the above elementary exchange relations it is possible to also exchange $U_1$ with $U_2$ between a pair of Lax matrices
\begin{equation}
L_{x_1, y_1}(V_1, U_1) L_{x_2, y_2}(V_2, U_2) =  L_{x_1', y_1'}(V_1, U_2)  L_{x_2', y_2'}(V_2, U_1).
\label{U_swap}
\end{equation}
This can be achieved by a composition of three elementary exchanges
\begin{equation}
(V_1, U_1), (V_2, U_2) \rightarrow (V_1, U_1), (U_2, V_2)  \rightarrow (V_1, U_2), (U_1, V_2) \rightarrow (V_1, U_2), (V_2, U_1).
\end{equation}
In an analogous fashion we can exchange $V_1$ with $V_2$
\begin{equation}
L_{x_1, y_1}(V_1, U_1) L_{x_2, y_2}(V_2, U_2) =  L_{x_1', y_1'}(V_2, U_1)  L_{x_2', y_2'}(V_1, U_2), \label{V_swap}
\end{equation}
via the following sequence of elementary exchanges
\begin{equation}
(V_1, U_1), (V_2, U_2) \rightarrow (U_1, V_1), (V_2, U_2)  \rightarrow (U_1, V_2), (V_1, U_2) \rightarrow (V_2, U_1), (V_1, U_2).
\end{equation}

\paragraph{Solving the zero-curvature condition.}
To obtain the two-body propagator $\Phi$ the discrete zero-curvature condition, Eq.~\eqref{eqn:ZC_abstract},
\begin{equation}
L_{x_1, y_1}(V_1, U_1) L_{x_2, y_2}(V_2, U_2) =  L_{x_1', y_1'}(V_2, U_2)  L_{x_2', y_2'}(V_1, U_1),
\label{ZC_Weyl}
\end{equation}
 can now be 
solved by composing a `$U$-exchange' \eqref{U_swap} with a `$V$-exchange' \eqref{V_swap}, in whichever order
\begin{equation}
(V_1, U_1), (V_2, U_2) \rightarrow (V_1, U_2), (V_2, U_1)  \rightarrow (V_2, U_2), (V_1, U_1) .
\end{equation}
 Their composition gives an explicit propagator $\Phi$ that solves the zero-curvature condition \eqref{ZC_Weyl}.
Because the elementary exchanges are canonical transformations, the resulting propagator is automatically canonical.
Explicit formulas for these compositions are rather cumbersome and we display them in \appref{app:weyl_propagator}.

\subsection{Two-body propagator}

In analogy to the isotropic case~\cite{krajnik2020kardar}, the discrete `Trotter' time-step $\tau \in \mathbb{R}$ enters the construction through the difference of the additive alternating (i.e. staggered) spectral parameters $\lambda^{\pm} \equiv \lambda \pm \tau/2$, carried by the light-cone Lax matrices, cf. Eq.~\eqref{eqn:ZC_abstract}. In terms of spectral parameter $u$ and $v$, this amounts to setting
\begin{equation}
u^{\pm} = \nu \, e^{\ii \varrho \, \lambda^{\pm} }, \qquad
v^{\pm} = \nu \, e^{-\ii \varrho \, \lambda^{\pm}},
\label{uv_Trotter_param}
\end{equation}
recalling that $\nu = e^{\varrho}$.
With the above prescription, the exchange relations (given by equations \eqref{U_swap} and \eqref{V_swap}) and the propagator
$\Phi$ simplify considerably. Explicit expressions can be found in \appref{app:weyl_propagator}).

\medskip

The elementary propagator $\Phi_{\tau,\varrho}$ can alternatively be written explicitly using
Sklyanin variables  $\mathcal{K}$, $\mathcal{S}^{\pm}$.
With aid of Eqs.~\eqref{Weyl} and some exercise, we find (where $w = e^{\ii \varrho \, \tau}$)
\begin{small}
\begin{align}
\left( \frac{\mathcal{K}'_1}{\mathcal{K}_1} \right)^2 &= 
 \frac{(\nu^2 + \nu^{-2})\left( \mathcal{K}_1 \mathcal{K}_2 +  \mathcal{K}_1^{-1} \mathcal{K}_2^{-1} \right) - 2\mathcal{K}_1 \mathcal{K}_2^{-1} - (w^2 + w^{-2}) \mathcal{K}_1^{-1} \mathcal{K}_2 + 4w \mathcal{S}^+_1 \mathcal{S}^-_2 + 4w^{-1}\mathcal{S}^-_1 \mathcal{S}^+_2}
{(\nu^2 + \nu^{-2})\left( \mathcal{K}_1 \mathcal{K}_2 +  \mathcal{K}_1^{-1} \mathcal{K}_2^{-1} \right) - 2\mathcal{K}_1^{-1} \mathcal{K}_2 - (w^2 + w^{-2}) \mathcal{K}_1 \mathcal{K}_2^{-1} + 4w \mathcal{S}^-_1 \mathcal{S}^+_2 + 4w^{-1}\mathcal{S}^+_1 \mathcal{S}^-_2},  \nonumber\\
\left( \frac{\mathcal{K}'_2}{\mathcal{K}_2} \right)^2 &= 
\frac{(\nu^2 + \nu^{-2})\left( \mathcal{K}_1 \mathcal{K}_2 +  \mathcal{K}_1^{-1} \mathcal{K}_2^{-1} \right) - 2\mathcal{K}_1^{-1} \mathcal{K}_2 - (w^2 + w^{-2}) \mathcal{K}_1 \mathcal{K}_2^{-1} + 4w \mathcal{S}^-_1 \mathcal{S}^+_2 + 4w^{-1}\mathcal{S}^+_1 \mathcal{S}^-_2}
 {(\nu^2 + \nu^{-2})\left( \mathcal{K}_1 \mathcal{K}_2 +  \mathcal{K}_1^{-1} \mathcal{K}_2^{-1} \right) - 2\mathcal{K}_1 \mathcal{K}_2^{-1} - (w^2 + w^{-2}) \mathcal{K}_1^{-1} \mathcal{K}_2 + 4w \mathcal{S}^+_1 \mathcal{S}^-_2 + 4w^{-1}\mathcal{S}^-_1 \mathcal{S}^+_2}, \nonumber \\
\begin{bmatrix}
(\mathcal{S}^{\pm}_1)^{\prime} \\
(\mathcal{S}^{\pm}_2)^{\prime}
\end{bmatrix}
&= \Omega(w^{\pm 1})
\begin{bmatrix}
\mathcal{S}^{\pm}_{1} \\
\mathcal{S}^{\pm}_{2}
\end{bmatrix},
\label{Sklyanin_eom}
\end{align}
\end{small}
with
\begin{equation}
\Omega(w) = \frac{1}{\mathcal{K}_1 \mathcal{K}_2 w - (\mathcal{K}_1 \mathcal{K}_2 w)^{-1}} \!
\begin{bmatrix}
\mathcal{K}^{\prime}_{1}\mathcal{K}_{2} - (\mathcal{K}^{\prime}_{1}\mathcal{K}_{2})^{-1} & \kern-1em
(\mathcal{K}^{\prime}_{1}/\mathcal{K}_{1})w -(\mathcal{K}_{1}/\mathcal{K}^{\prime}_{1})w^{-1} \\
(\mathcal{K}^{\prime}_{2}/\mathcal{K}_{2})w -(\mathcal{K}_{2}/\mathcal{K}^{\prime}_{2})w^{-1} & \kern-1em
\mathcal{K}_{1}\mathcal{K}^{\prime}_{2} - (\mathcal{K}_{1}\mathcal{K}_{2}^{\prime})^{-1}
\end{bmatrix}.
\end{equation}
We remind the reader that in the easy-axis regime $\mathcal K \in \mathbb{R}_{+}$, whereas in the easy-plane regime
$\ani \, S^{3} = \textrm{arg}(\mathcal{K}) \in [-\pi/2, \pi/2]$.
Therefore, there is no ambiguity in Eqs.~\eqref{Sklyanin_eom} when taking the square roots to determine $\mathcal{K}'_\ell$.

\paragraph*{Properties.}
There are a number of important properties worth highlighting:
\begin{itemize}
\item In the continuous-time limt $\tau \to 0$, the propagator $\Phi_{\tau,\varrho}$ reduces to the identity map.
\item The mapping \eqref{Sklyanin_eom} manifestly preserves the product $\mathcal{K}_1 \mathcal{K}_2$, implying that the local propagator
conserves the third component of the spin,
\begin{equation}
{S}_1^{3} + {S}_{2}^{3} = ({S}_{1}^{3})' + ({S}_{2}^{3})'.
\label{two_body_conservation}
\end{equation}
This local $U(1)$ symmetry directly implies conservation of $S^{3}_{\rm tot}\equiv \sum_{\ell=1}^{L}S^{3}_{\ell}$ under the full propagator
$\Phi^{\rm full}$ which, in the Hamiltonian limit, $\tau \to 0$, becomes the Noether charge of
the uniaxially anisotropic Landau--Lifshitz model.
\item In the easy-axis regime, the local propagator \eqref{Sklyanin_eom} is a \emph{periodic} function of interaction $\varrho$,
with period $2\pi/\tau$.
\item The two-body propagator $\Phi_{\tau,\varrho}$ is invariant under the transformation
\begin{equation}
\varrho \rightarrow - \varrho, \quad \tau \rightarrow -\tau, \qquad \vec{S}_{1,2} \rightarrow -\vec{S}_{1,2}.
\label{symmetries}
\end{equation}
It is therefore sufficient to consider only the interval $\varrho \in [0, \pi/\tau]$.
\item In the \emph{easy-axis regime}, at $\varrho = \pi/\tau$ (equiv. $w = -1$) the dynamics trivializes to the identity for all $\tau$,
\begin{equation}
\lim_{\varrho \to \pi/\tau}\Phi_{\tau,\varrho} (\vec{S}_1, \vec{S}_2) = (\vec{S}_1, \vec{S}_2).
\label{easy_axis_id}
\end{equation}
\item In the \emph{easy-plane regime}, the dynamics is \emph{not} periodic in $\varrho$, but is instead confined to the interval
$\ani \in [-\pi/2,\pi/2]$. By virtue of Eqs.~\eqref{symmetries}, it is sufficient to consider only the range $\ani \in [0, \pi/2]$.
\item In the easy-plane regime, in the limit $\tau \rightarrow \infty$ with $\ani$ fixed (i.e. $|w| \rightarrow \infty$),
$\Phi_{\tau,\ii\ani}$ reduces to a permutation
\begin{equation}
\lim_{\tau\to \infty}\Phi_{\tau,\ii\ani} (\vec{S}_1, \vec{S}_2) = (\vec{S}_2, \vec{S}_1).
\label{easy_plane_swap}
\end{equation}
\item The two-body propagator $\Phi$ is a \emph{classical Yang--Baxter map} \cite{Veselov03,Hietarinta_book,BS18,Tsuboi18},
obeying the braided Yang--Baxter relation on the Cartesian product $\mathcal{M}^{\times 3}_{1}$ (suppressing the anisotropy parameter)
\begin{equation}
\Phi^{(1,2)}_{\lambda_{2}-\lambda_{3}}\circ \Phi^{(2,3)}_{\lambda_{1}-\lambda_{3}}\circ \Phi^{(1,2)}_{\lambda_{1}-\lambda_{2}} =
\Phi^{(2,3)}_{\lambda_{1}-\lambda_{2}}\circ \Phi^{(1,2)}_{\lambda_{1}-\lambda_{3}}\circ \Phi^{(2,3)}_{\lambda_{2}-\lambda_{3}},
\end{equation}
where $\Phi^{(i,j)}$ signifies that $\Phi$ operates non-identically on $i$-th and $j$-th copy of $\mathcal{M}^{\times 3}_{1}$.
\end{itemize} 

\paragraph{Discrete Noether current.}

The lattice Landau--Lifshitz model, cf. Eq.~\eqref{LLL_Hamiltonian}, is invariant under global rotations about the third axis. This
imples, by virtue of the Noether theorem, a globally conserved Noether charge (magnetization) in the system.
We have just explained, cf. Eq.~\eqref{two_body_conservation}, that the same property holds even in the discrete-time setting.

\begin{figure}[htb]
\centering
\begin{tikzpicture}[scale=1.5]
\tikzstyle{curarrow}=[line width = 1mm, draw=teal!50,-triangle 45,postaction={draw, line width = 2.5mm, shorten >=3mm, -}]


\begin{scope}[rotate=45]
\draw[fill=black!5,very thick, drop shadow] (-2,2) rectangle (0,4);
\draw[fill=black!5,very thick, drop shadow] (0,0) rectangle (2,2);
\draw[fill=black!15,very thick] (0,2) rectangle (2,4);

\draw[dashed] (-1,1) rectangle (1,3);


\node[single arrow, rounded corners = 2pt, fill=pink!50, draw, drop shadow, align=center, xshift = 0.1 cm, yshift = 0 cm,
minimum height=1.1cm, minimum width = 0.5cm, shape border rotate = 180] at (-1.5,2.5) {\tiny $\!\!j^{2t}_{2\ell}$};
\node[single arrow, rounded corners = 2pt, fill=pink!50, draw, drop shadow, align=center, xshift = 0.1 cm, yshift = 0 cm,
minimum height=1.1cm, minimum width = 0.5cm, shape border rotate = 180] at (0.5,2.5) {\tiny $\!\!j^{2t+1}_{2\ell+1}$};

\foreach \i in {1,...,2} {
	\draw[gray,decoration={markings,mark=at position 0.65 with
    		{\arrow[scale=1.5,>=stealth']{latex}}},postaction={decorate}] (-3+2*\i,5-2*\i) -- (-3+2*\i+1,5-2*\i);
    	\draw[gray,decoration={markings,mark=at position 0.55 with
    		{\arrow[scale=1.5,>=stealth']{latex}}},postaction={decorate}] (-3+2*\i-1,5-2*\i) -- (-3+2*\i,5-2*\i);
	\draw[gray,decoration={markings,mark=at position 0.55 with
    		{\arrow[scale=1.5,>=stealth']{latex}}},postaction={decorate}] (-3+2*\i,5-2*\i-1) -- (-3+2*\i,5-2*\i);
	\draw[gray,decoration={markings,mark=at position 0.6 with
    		{\arrow[scale=1.5,>=stealth']{latex}}},postaction={decorate}] (-3+2*\i,5-2*\i) -- (-3+2*\i,5-2*\i+1);

	\node at (-3+2*\i,5-2*\i) [fill=red!50,inner sep=0pt,minimum size=24pt,draw,rounded corners, drop shadow] (P1i) {$\Phi$};
}

\foreach \i in {1,...,1} {

	\draw[gray,decoration={markings,mark=at position 0.65 with
    		{\arrow[scale=1.5,>=stealth']{latex}}},postaction={decorate}] (-1+2*\i,5-2*\i) -- (-1+2*\i+1,5-2*\i);
    	\draw[gray,decoration={markings,mark=at position 0.55 with
    		{\arrow[scale=1.5,>=stealth']{latex}}},postaction={decorate}] (-1+2*\i-1,5-2*\i) -- (-1+2*\i,5-2*\i);
	\draw[gray,decoration={markings,mark=at position 0.55 with
    		{\arrow[scale=1.5,>=stealth']{latex}}},postaction={decorate}] (-1+2*\i,5-2*\i-1) -- (-1+2*\i,5-2*\i);
	\draw[gray,decoration={markings,mark=at position 0.6 with
    		{\arrow[scale=1.5,>=stealth']{latex}}},postaction={decorate}] (-1+2*\i,5-2*\i) -- (-1+2*\i,5-2*\i+1);

	\node at (-1+2*\i,5-2*\i) [fill=red!50,inner sep=0pt,minimum size=24pt,draw,rounded corners, drop shadow] (P1i) {$\Phi$};
}

\foreach \i in {1,2} {

	\node at (-3+\i+3,4-\i-1) [fill=teal!10,inner sep=0pt,minimum size=24pt,draw,circle,drop shadow] (M2i) {\scriptsize{$q^{2t+1}_{2\ell+\i}$}};
}
\foreach \i in {1,2} {

	\node at (-3+\i+2,4-\i-2) [fill=teal!10,inner sep=0pt,minimum size=24pt,draw,circle,drop shadow] (M1i) {\scriptsize{$q^{2t}_{2\ell+\i}$}};
}
%
%

\node at (-3+2+2,4-2+2) [fill=teal!30,inner sep=0pt,minimum size=24pt,draw,circle,drop shadow] (M32) {\scriptsize{$q^{2t+2}_{2\ell}$}};
\node at (-3+3+2,4-3+2) [fill=teal!10,inner sep=0pt,minimum size=24pt,draw,circle,drop shadow] (M33) {\scriptsize{$q^{2t+2}_{2\ell+1}$}};

\node at (-3-1+3,4+1-1) [fill=teal!10,inner sep=0pt,minimum size=24pt,draw,circle,drop shadow] (M2i) {\scriptsize{$q^{2t+1}_{2\ell-1}$}};
\node at (-3+3,4-1) [fill=teal!10,inner sep=0pt,minimum size=24pt,draw,circle,drop shadow] (M2i) {\scriptsize{$q^{2t+1}_{2\ell}$}};

\node at (-3-1+2,4+1-2) [fill=teal!10,inner sep=0pt,minimum size=24pt,draw,circle,drop shadow] (M1i) {\scriptsize{$q^{2t}_{2\ell-1}$}};
\node at (-3+2,4-2) [fill=teal!30,inner sep=0pt,minimum size=24pt,draw,circle,drop shadow] (M12) {\scriptsize{$q^{2t}_{2\ell}$}};

\end{scope}

\end{tikzpicture}
\caption{Discrete Noether current, shown at an even site $2\ell$: the time-increment $t\to t+2$ of charge density,
$\big(q^{t+2}_{2\ell}-q^{t}_{2\ell}\big)/\tau$ equals the difference of two nearby current densities at consecutive times,
$j^{2t+1}_{2\ell+1}-j^{2t}_{2\ell}$. Current densities $j^{t}_{\ell}$, depicted by pink arrows,
can be compactly expressed as the difference of charge densities at consecutive times,
$j^{t}_{\ell}=(q^{t+1}_{\ell}-q^{t}_{\ell})/\tau$.}
\label{fig:current}
\end{figure}
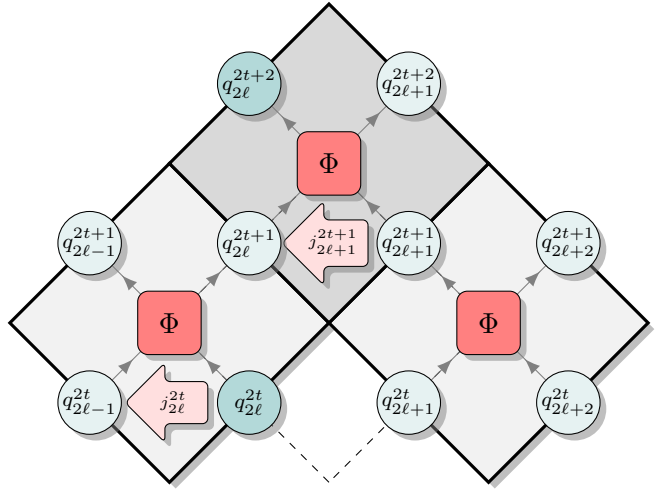

By virtue of the global $U(1)$ symmetry, the dynamics admits a conserved Noether current. Its spatial component, i.e.
the `charge density', is given by the projection of ${\bf S}$ onto the third axis, $q = S^{3}$. 
The associated temporal component is the `current density' which we subsequently denote by $j$.
Owing to the even-odd staggered structure of the full propagator $\Phi^{\rm full}$,
the discrete continuity equations at even and odd sites are of the form 
\begin{equation}
\frac{1}{\tau}\left(q_{2\ell}^{2t+2} - q_{2\ell}^{2t}\right) + j_{2\ell+1}^{2t+1} - j_{2\ell}^{2t} = 0,  \qquad
\frac{1}{\tau}\left( q_{2\ell+1}^{2t+2}  - q_{2\ell+1}^{2t} \right) + j_{2\ell+2}^{2t} - j_{2\ell+1}^{2t+1} = 0,
\label{discrete_continuity}
\end{equation}
respectively. While the charge densities $q^{t}_{\ell}\equiv q(\ell,t)$ are ultralocal, namely they sit at site $\ell$ and time $t$,
the associated current density corresponds to their forward differences~\footnote{We note that the current density is only defined up to a total difference. We fix this gauge freedom by demanding that the current vanishes when when both input spins are equal.}
\begin{equation}
j_{\ell}^{t} \equiv j(\ell,t) = \frac{1}{\tau}\big(q(\ell,t+1) - q(\ell,t)\big),
\label{discrete_j}
\end{equation}
which can be immediately verified by plugging the expression into the discrete continuity equations \eqref{discrete_continuity} and
using the conservation property \eqref{two_body_conservation}.
Notice that $q^{t+1}_{\ell}$ is a function of two adjacent charge densities at the previous time slice, namely
$q^{t}_{\ell}$ and $q^{t}_{\ell \pm 1}$, cf. \figref{fig:current}, that enter the local propagator $\Phi$.

\paragraph{Isotropic, continuous time and field-theory limits.}
The two-parametric dynamical map  $\Phi^{\rm full}_{\tau,\varrho}$ admits various important limits:
(i) the limit of vanishing anisotropy $\varrho \to 0$, (ii) the continuous-time limit $\tau \to 0$ and (iii)
the continuous space-time (i.e. field-theory) limit by additionally taking the long wavelength limit.
The resulting models, outlined in \appref{app:reductions}, can be all regarded as members of the `Landau-Lifshitz family'.

\section{Magnetization transport}
\label{sec:transport}

We demonstrate the utility of the constructed dynamical system by studying magnetization transport in thermal equilibrium.
We focus on numerical computation of the linear transport coefficients that quantify magnetization transport on a large spatio-temporal
(i.e. hydrodynamic) scale, namely the spin Drude weight and the spin diffusion constant.
After introducing the equilibrium measure and transport coefficients we proceed by systematically analyzing the
easy-axis and the easy-plane regimes.

\subsection{Transport coefficients}

\paragraph{Equilibrium ensemble.}
We are interested in the dynamical response in our model at finite magnetization density.
To this end, we introduce a one-parametric family of local measures $\rho_{1}(\mu)$ on $\mathcal{M}_{1}$,
\begin{equation}
\rho_{1}(\vec{S}) = \frac{1}{4\pi} \frac{\mu}{\sinh \mu} e^{\mu \, S^3},
\end{equation}
which can be interpreted as the grand-canonical measure with
chemical potential $\mu$.~\footnote{Such a one-body measure maximizes the Shannon (or Gibbs) entropy of a state, that is
$-\int \dd \Omega \, \rho_1(\vec{S}) \log \rho_1(\vec{S})$ where $\dd \Omega$ is a uniform measure on $S^{2}$ with a prescribed average $\langle S^{3}\rangle$ set by a Lagrange multiplier $\mu$. In the continuous-time (i.e. Hamiltonian) limit,
$\tau \to 0$, $\rho_{1}(\vec{S})$ is the grand-canonical Gibbs measure in the limit of infinite temperature.}
We accordingly define a separable \emph{stationary} measure
$\rho_{L}\equiv \rho_{L}\left(\{\vec{S}_{\ell}\}_{\ell = 1}^{L}\right)$ on the full phase space $\mathcal{M}_{L}$,
\begin{equation}
\rho_{L} = \prod_{\ell=1}^{L}\rho_{1}(\vec{S}_{\ell}).
\label{many_body_measure}
\end{equation}
with chemical potential $\mu$ enforcing a non-zero average of magnetization.
Invariance of $\rho_{L}$ under the full dynamics $\Phi^{\rm full}$ follows from the fact that the quadratic Sklyanin
bracket is preserved by the action of $\Phi$ (see Ref.~\cite{krajnik2020integrable} for an analogous formal proof).

We now pass over to the thermodynamic limit by sending $L\to \infty$. The grand-canonical free energy per site is given by
\begin{equation}
f(\mu) = -\log \mathcal{Z}_{1}(\mu),
\end{equation}
where
\begin{equation}
\mathcal{Z}_{1}(\mu)=\int_{S^{2}}\dd \Omega\,\rho_{1}(\vec{S}) = \frac{4\pi \sinh{\mu}}{\mu},
\end{equation}
represents the onsite grand-canonical partition function, with $\dd \Omega$ denoting the volume element on $S^{2}$.
Writting $\dd \Omega^{\rm full} \equiv \prod_{\ell=1}^{L} \dd \Omega(\vec{S}_{\ell})$, the corresponding thermodynamic
average of any local observable $\mathcal{O}$ is therefore computed as
\begin{equation}
\langle \mathcal{O} \rangle_{\mu} = \lim_{L\to \infty}\big[\mathcal{Z}^{L}_{1}(\mu)\big]^{-1}
\int_{\mathcal{M}_{L}}\dd \Omega^{\rm full}\,\rho_{L}\,\mathcal{O}.
\end{equation}
For example, the average of magnetization density reads 
\begin{equation}
\langle S^{3} \rangle_{\mu} = \coth{\mu}-1/\mu.
\end{equation}
In the following we accordingly take $L\to \infty$ and operate in the thermodynamic setting.

\paragraph*{Conductivity.}
In the linear-reponse regime, transport coefficients are encoded in the real part of the frequency-dependent conductivity $\sigma(\omega)$,
which is in general decomposed as
\begin{equation}
\sigma(\omega) = \pi\mathcal{D}\,\delta(\omega) + \sigma_{\rm reg}(\omega).
\end{equation}
The $\delta$-peak owes its existence to the presence of long-lived excitations which ballistically transport the charges through the system,
characterized by a coefficient $\mathcal{D}$ called the Drude weight. The d.c. component of the `regular' part, that is
$\sigma_{\rm reg}(0)$, characterizes transport on sub-ballistic scales. Provided $\sigma_{\rm reg}(0)$ is finite, it
can be associated to  the spin diffusion constant defined as
\begin{equation}
D = \chi^{-1} \lim_{\omega \to 0}\sigma_{\rm reg}(\omega),
\end{equation}
where $\chi$ denotes static spin susceptibility, $\chi(\mu) = \partial^{2} f(\mu)/\partial^{2} \mu = 1 + \mu^{-2} - \coth^{2}\mu$.
In integrable \emph{interacting} systems, the diffusion constant quantifies the diffusive spreading which quasiparticle excitations exhibit
upon ellastic collisions (see e.g. Refs.~\cite{Doyon_lectures,diffusion_review}).
As customary, we define the dynamical structure factor
\begin{equation}
S(\ell,t) = \langle q(\ell,t)\,q(0,0) \rangle^{c}_{\mu},
\label{dsf}
\end{equation}
with a sum rule $\sum_{\ell \in \mathbb{Z}}S(\ell,t) = \chi$. The spin Drude weight governs the time-asymptotic growth
of its second moment
\begin{equation}
\sum_{\ell \in \mathbb{Z}} \ell^{2} \, S(\ell,t) \simeq \mathcal{D}\,t^{2}+ \mathcal{O}(t).
\label{second_moment}
\end{equation}
where the subleading correction, which equals $2\chi\,D$, stores information about the spin diffusion constant $D$.
We note that numerically extracting $D$ from the $\mathcal{O}(t)$ broadening of the second moment \eqref{second_moment} has not
proven particularly reliable.

\medskip

We shall employ an alternative approach and define the (linear) transport coefficients through the Kubo formula.
The extensive magnetization current $J$ at physical times $2t$ follows directly from the discrete continuity equation \eqref{discrete_continuity} 
and is given by
\begin{equation}
J(2t) = \sum_{\ell \in \mathbb{Z}} \big(j(2\ell,2t) + j(2\ell+1,2t+1)\big),
\end{equation}
The spin Drude weight corresponds to the asymptotic value of the current autocorrelation function
\begin{equation}
\mathcal{D} = \lim_{t \rightarrow \infty} \frac{1}{t} \sum_{t'=0}^t  \, \langle J(2t') j(0,0) \rangle^{c}_{\mu},
\label{Drude_def}
\end{equation}
whereas the spin diffusion constant is the integrated current autocorrelator with the asymptotic value subtracted
\begin{equation}
D = \chi^{-1} \lim_{t \rightarrow \infty} \tau \sum_{t'=0}^t \, \left(\langle J(2t') j(0,0)\rangle^{c}_{\mu}  - \mathcal{D} \right).
\label{diff_def}
\end{equation}
Lastly, we introduce the dynamical exponent $\alpha$ that characterizes the late-time decay of the dynamical spin structure factor
\begin{equation}
S(0,t) \sim t^{-\alpha}.
\label{finite_time_exponent}
\end{equation}
Ballistic spreading corresponds to exponent $\alpha=1$. On the other hand, we can speak of normal diffusion when
(i) $\alpha=1/2$ and (ii) the dynamical structure factor converges asymptotically to a Gaussian stationary scaling profile,
\begin{equation}
S(\ell,t)\simeq \frac{\chi}{\sqrt{2\pi D t}}e^{-\ell^{2}/2Dt}.
\end{equation}

\begin{figure}[h]
\centering
\includegraphics[width=\textwidth]{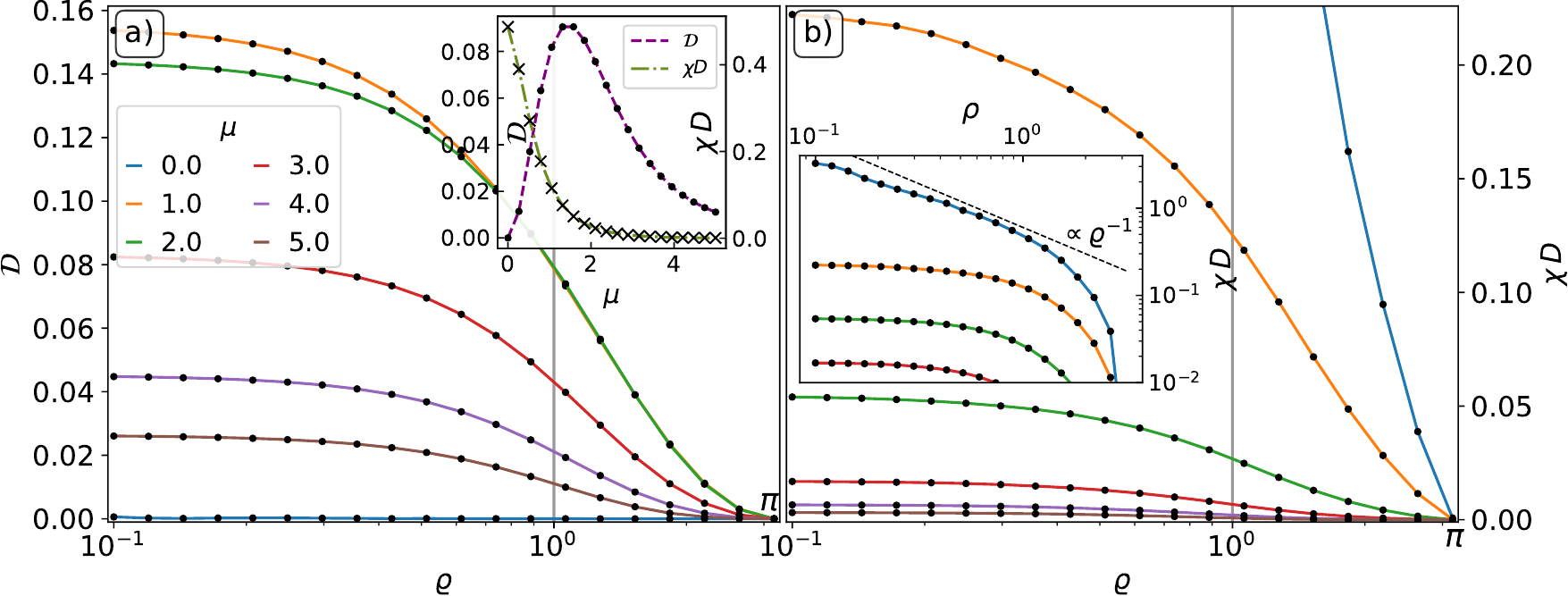}
\caption{(a) Spin Drude weight $\mathcal{D}$ and (b) diffusion constant $D$ in the discrete Landau--Lifshitz model,
cf. Eq.~\ref{eqn:st}, in the easy-axis regime, depending on the chemical potential $\mu$ and the anisotropy $\varrho$.
(a, inset) Drude weight (purple) and diffusion constant (olive green) as functions of $\mu$ at $\varrho = 1$ (vertical grey line). 
(b, inset) Divergence of $D$ at zero average magnetization, $\mu = 0$ (blue curve),
in the isotropic limit $\varrho \rightarrow 0$, compared to $\sim 1/\varrho$ power-law scaling (black dashed line).
Simulation parameters: $\tau = 1$, $L=2^{12}$, averaged over $10^{5}$ initial configurations.}
\label{fig:easy_axis1}
\end{figure}

\subsection{Numerical simulations: easy-axis regime}

\begin{figure}[h]
\centering
\includegraphics[width=\textwidth]{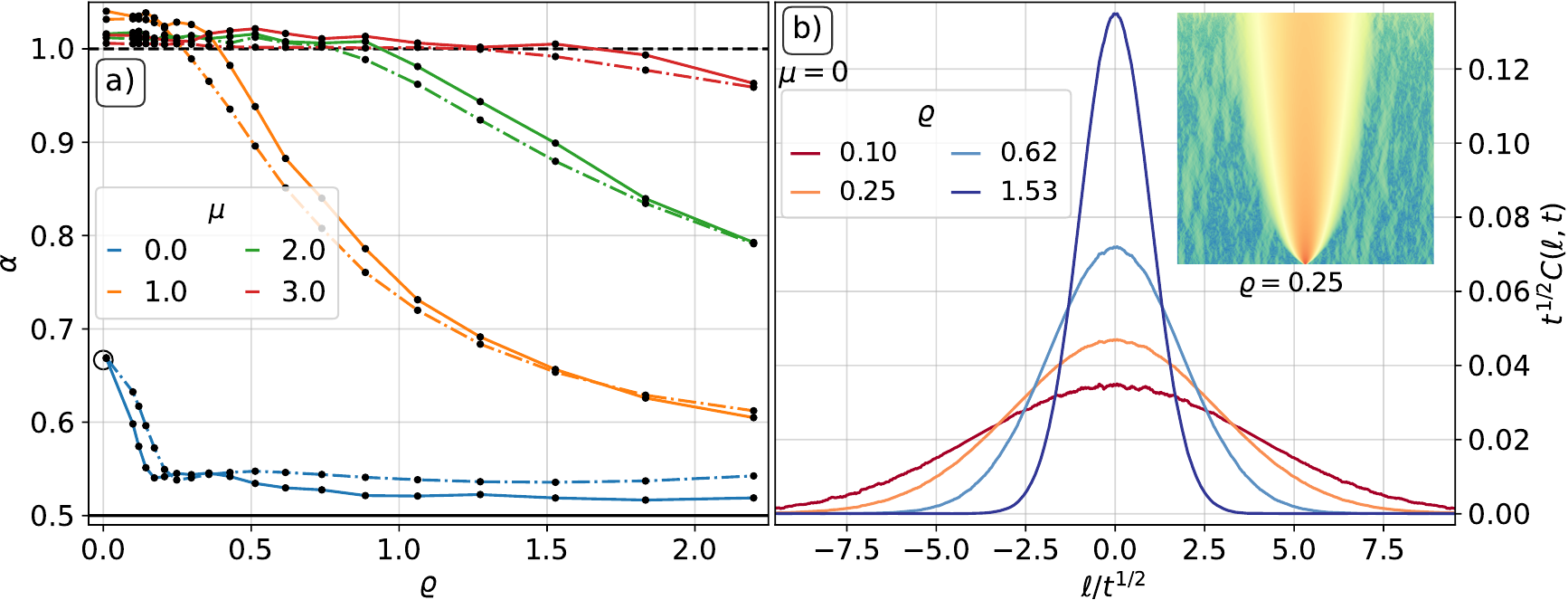}
\caption{(a) Finite-time dynamical exponents $\alpha$ ($t \in [100, 300]$ dashed-dotted lines, $t \in [300,1000]$ solid lines) depending on the anisotropy $\varrho$ at four values of the chemical potentials $\mu$. Magnetized ensembles ($\mu > 0$) exhibit ballistic transport with
exponent $\alpha = 1$. The sub-leading diffusive correction increases with $\varrho$ and overwhelms the ballistic contribution at finite times.
In the non-magnetized sector at $\mu = 0$, the dynamical exponent is approximately $\alpha \approx 2/3$ (black circle) when approaching
$\varrho \rightarrow 0$, consistent with KPZ-type superdiffusion at the isotropic point $\varrho = 0$. At large values of $\varrho$, the dynamical exponent approaches $\alpha \approx 1/2$ which is indicative of normal diffusion. (b) Rescaled diffusive stationary scaling profiles of the dynamical
structure factors $t^{1/2}\,C(\ell,t)$ at $\mu = 0$ and different values of $\varrho$.
Simulation parameters: $\tau = 1$, $L=2^{12}$, averaged over $10^{5}$ initial configurations.}
\label{fig:easy_axis2}
\end{figure}

We first focus our analysis on the spin Drude weight and diffusion in the easy-axis regime. The results are shown
in \figref{fig:easy_axis1}. The asymptotic stationary profiles and the numerically estimated dynamical exponents are shown
in \figref{fig:easy_axis2}.

\paragraph*{Drude weight.}
The spin Drude weight is finite for all anisotropies $\varrho$ and arbitrary finite magnetization density $\mu>0$. In the unmagnetized ensemble ($\mu=0$),
corresponding to a uniform measure $\rho_{1}=1/{\rm Vol}(S^{2})=1/4\pi$, the Drude weight vanishes. This point is distinguished by the discrete
$\mathbb{Z}_{2}$ symmetry, corresponding to the global reversal of all spins, $S_\ell^{\pm}\mapsto S_\ell^{\mp}$, $S_\ell^{z}\mapsto -S_\ell^{z}$,
which is a canonical transformation, i.e. preserving the Poisson bracket.
We find, moreover, that both transport coefficients vanish when $\varrho \rightarrow \pi/\tau$, consistent with the dynamics becoming trivial in 
that limit, cf. Eq.~\eqref{easy_axis_id}.

\begin{figure}[h!]
\centering
\includegraphics[width=\textwidth]{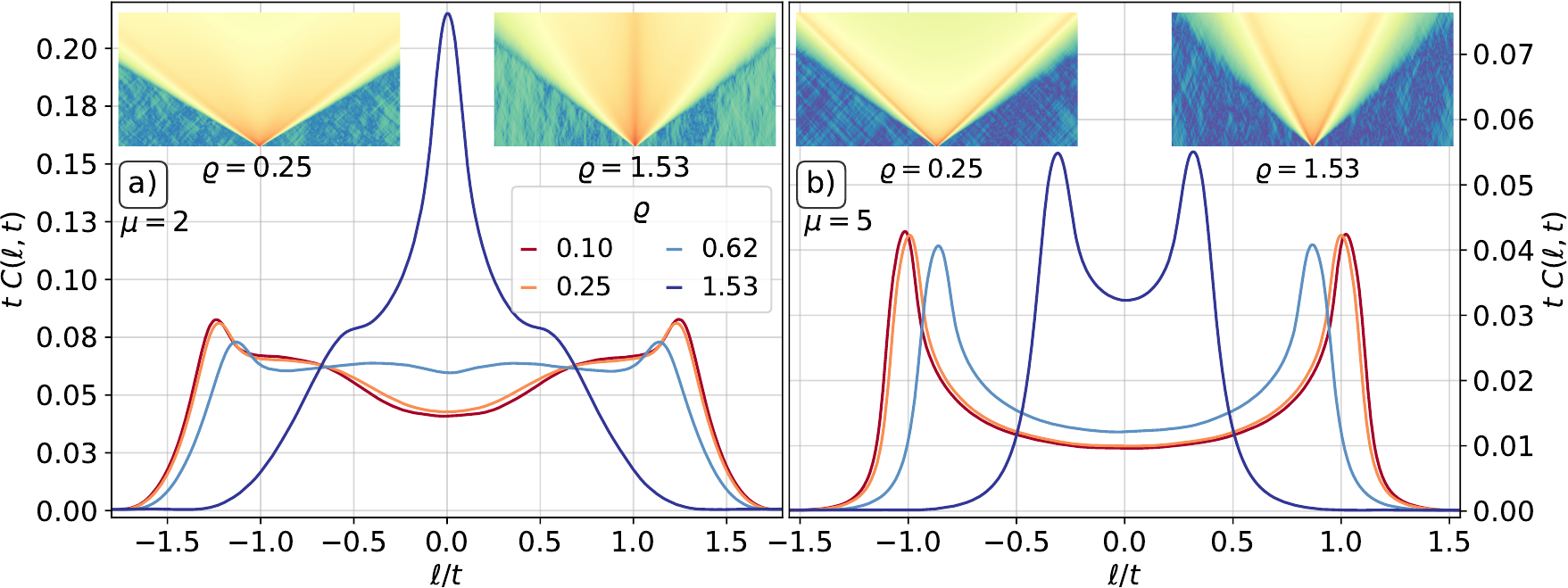}
\caption{Ballistic stationary scaling profiles of dynamical structure factors $S(\ell,t)$ (shown in insets) computed at $\mu = 2, 5$.
Simulation parameters: $\tau = 1$,$L=2^{12}$, averaged over $10^{5}$ initial configurations.}
\label{fig:easy_axis3}
\end{figure}

Fixing anisotropy to $\varrho=1$, we find a non-monotonic dependence of $\mathcal{D}$  on  the chemical potential $\mu$,
see the inset plot in \figref{fig:easy_axis1}, which is consistent with the picture of dressed quasiparticles~\cite{diffusion_review,superdiffusion_review}: in an unpolarized ensemble, the dressed magnetization vanishes and so does the spin Drude weight. For small $\mu$, the quasiparticles behave paramagnetically, with dressed magnetization behaving as $\sim \mu$.
For strong polarizations, $\mu \to \infty$, one approaches a $\mathbb{Z}_{2}$-degenerate quasiparticle pseudovacuum where
the quasiparticles become effectively free. In this regime, however, the quasiparticle density vanishes and the spin Drude weight
drops to zero.

\paragraph*{Diffusion constant.}
The spin diffusion constant $D$ is finite for all $\mu > 0$. We find that it decreases monotonically with increasing $\mu$. At any given
anisotropy $\varrho$, $D$ grows upon lowering $\mu$, and attains its maximal value at $\mu = 0$. There, upon approaching the isotropic point,
$\varrho \to 0$, the diffusion constant $D$ diverges in (approximately) algebraic fashion, $D\sim 1/\varrho^{\varkappa_{\rm ea}}$,
as shown in \figref{fig:easy_axis1} (right inset). This conforms with a theoretical expectation $\varkappa_{\rm ea} = 1$ based on
the known behavior in the quantum Heisenberg chain \cite{GV19}.~\footnote{In the gapped phase of the Heisenberg spin-$1/2$ chain
with anisotropy $\Delta=\cosh{(\eta)}$, the spin diffusion constant diverges as $\sim 1/\eta$ as $\eta \to 0$.
In the semiclassical limit, easy-axis anisotropy $\varrho$ is proportional to $\eta$, see \appref{app:semiclassical}.} 
A reliable estimation of $\varkappa_{\rm ea}$ from the numerical data, as shown in \figref{fig:easy_axis1} (b, inset, dashed black line), is however rather difficult due to a divergent relaxation timescale in the proximity of the isotropic point.
Singularity of the spin diffusion constant signals the onset of \emph{superdiffusion} \cite{MarkoKPZ} which has recently been under intense scrutiny \cite{LjubotinaNature17,Ilievski18,LjubotinaPRL19,NMKI19,DasKPZ,krajnik2020kardar,Vir20,krajnik2020integrable}.
There exists ample numerical evidence \cite{LjubotinaPRL19,DupontMoore19,DasKPZ,krajnik2020kardar,krajnik2020integrable,DGIV20,superuniversality}
that this phenomenon belongs to the universality class of the Kardar--Parisi--Zhang (KPZ) equation \cite{KPZ86}.
In \figref{fig:easy_axis2}, we numerically corroborate the predicted anomalous dynamical exponent exponent $\alpha=2/3$.

\begin{figure}[h]
\centering
\includegraphics[width=\textwidth]{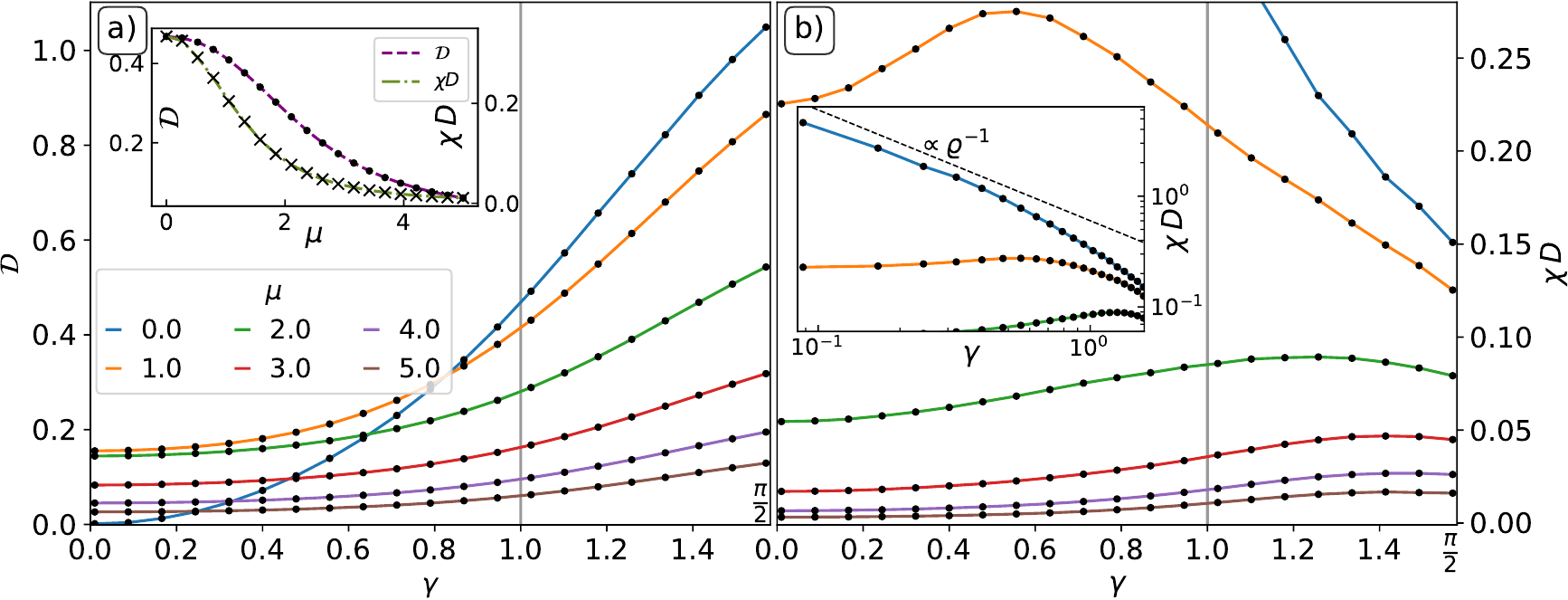}
\caption{(a) Spin Drude weight $\mathcal{D}$ and (b) and diffusion constant $D$ in the discrete Landau--Lifshitz model,
cf. Eq.~\eqref{eqn:st}, in the easy plane-regime, depending on chemical potential $\mu$ and anisotropy $\ani$. 
(a, inset) Drude weight (purple) and diffusion constant (olive green) as functions of $\mu$ at $\ani = 1$ (vertical grey line).
(b, inset) Divergence of $D$ at zero average magnetization, $\mu = 0$ (blue curve), in the isotropic limit
$\ani \rightarrow 0$, compared to $\sim 1/\varrho$ power-law scaling (black dashed line).
Simulation parameters: $\tau = 1$ , $L=2^{12}$, averaged over $10^{5}$ initial configurations.}
\label{fig:easy_plane1}
\end{figure}

\begin{figure}[h]
\centering
\includegraphics[width=\textwidth]{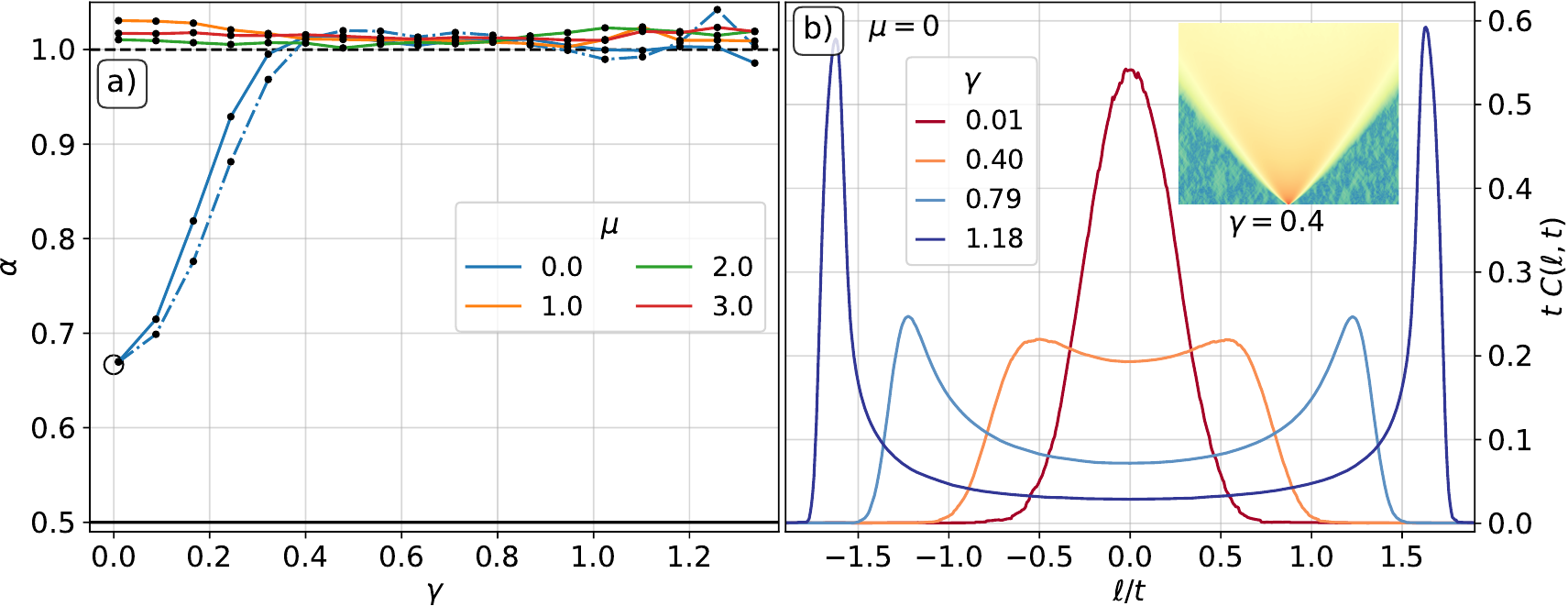}
\caption{(a) Finite-time dynamical exponents $\alpha$ (full blue line for $t \in [300, 1000]$, blue dashed-dotted line for $t\in  [100, 300]$) as functions of anisotropy for different values of chemical potentials $\mu$. At all $\mu > 0$, we find ballistic transport ($\alpha = 1$).
At $\mu = 0$ and $\ani \rightarrow 0$, the dynamical exponent crosses over to superdiffusion with $\alpha = 2/3$ (black circle).
(b) Rescaled ballistic stationary profiles of the dynamical structure factor $t\,C(\ell,t)$ at $\mu = 0$.}
\label{fig:easy_plane20}
\end{figure}

The transient effects can be discerned by plotting the dynamical structure factors (the equal-time profiles of dynamical correlation functions of the spin density), as shown in \figref{fig:easy_axis2}. While the dynamics is ballistic for all values of $\varrho>0$ and $\mu>0$, the finite-time dynamical exponent estimated based on \eqref{finite_time_exponent} is found to be ballistic, $\alpha=1$, only below a threshold value of $\varrho$ (which increases with increasing $\mu$). This is due to the fact that the value of $\mathcal{D}$ is small 
in comparison with $D$. By contrast, at $\mu = 0$, the estimated exponent is approximately diffusive, $\alpha \approx 1/2$
for $\varrho \gtrsim 1$, while when approaching $\varrho \to 0$ it starts to drift towards the expected superdiffusive exponent $\alpha=2/3$.
At $\mu = 0$, the stationary cross-sections shown in  \figref{fig:easy_axis2} (b) are single-peaked, consistent
with the absence of ballistic transport. As $\varrho \rightarrow 0$, the broadening of the central peak is compatible with
the aforementioned divergence of the diffusion constant.

\subsection{Numerical simulations: easy-plane regime}

\begin{figure}[h!]
\centering
\includegraphics[width=\textwidth]{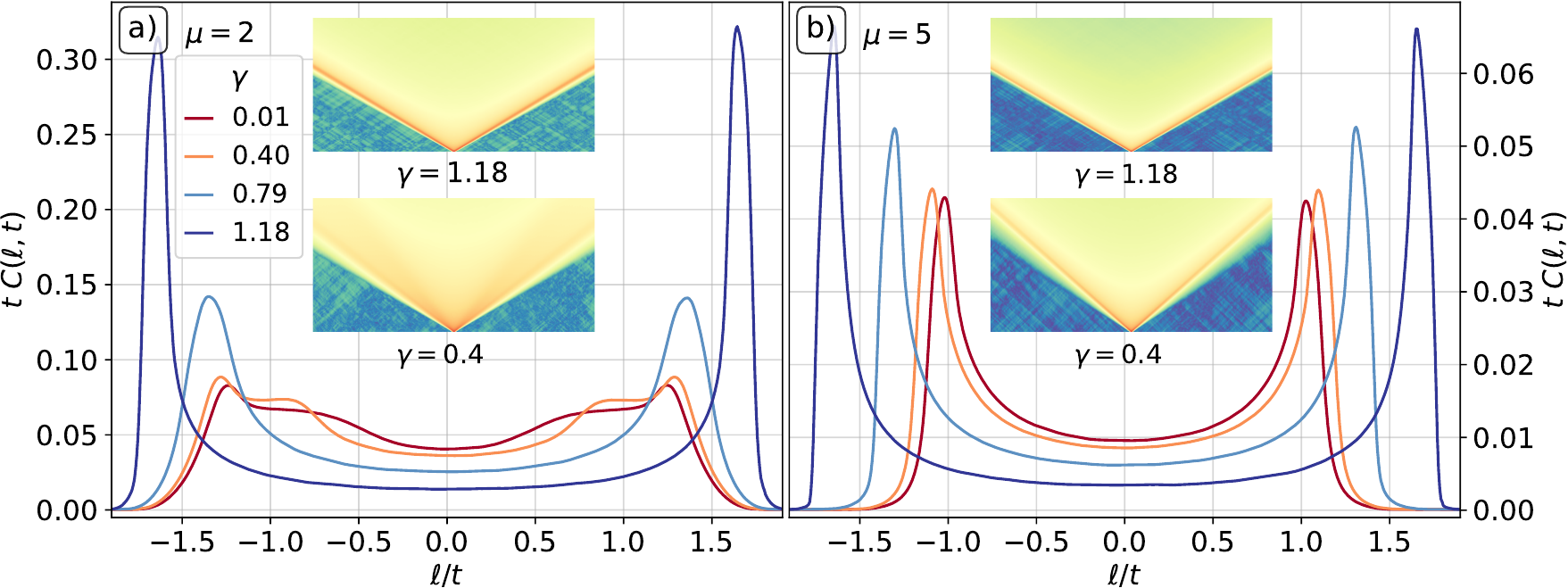}
\caption{Rescaled ballistic stationary profiles of the dynamical structure factor $t\,C(\ell,t)$ for (a) $\mu = 2$ and (b) $\mu = 5$,
with the corresponding space-time dependence $C(\ell,t)$ (inset plots).
Simulation parameters: $\tau = 1$, $L=2^{12}$, averaged over $10^{5}$ initial configurations.}
\label{fig:easy_plane21}
\end{figure}

We now repeat the above numerical analysis for the easy-plane regime. The spin Drude weights and diffusion constants are displayed
in \figref{fig:easy_plane1}, while the asymptotic stationary profiles of the dynamical structure factors, alongside the
finite-time dynamical exponents, are shown in \figref{fig:easy_plane20} and \figref{fig:easy_plane21}.

\paragraph*{Drude weight.}
In the easy-plane regime, the spin Drude weight $\mathcal{D}$ takes a \emph{non-zero} value for all values of $\mu$ and $\ani$,
except at the isotropic point at zero magnetization ($\mu = \ani = 0$). Upon lowering $\ani$, the spin Drude weight $\mathcal{D}$ now monotonously \emph{decreases},
for any fixed $\mu$, to a limiting value at $\ani=0$ which is non-zero for $\mu\neq 0$.

It is worth highlighting that $\mathcal{D}$ does not vanish even in a non-magnetized ensemble, in stark contrast with the easy-axis regime.
Altough this is perfectly aligned with previous numerical studies of the lattice Landau--Lifshitz model \cite{PZ13,DasKPZ},
a proper theoretical explanation is currently still lacking. At this juncture, we can offer two complementary interpretations:
\begin{enumerate}[label=(\alph*)]
\item from the quasiparticle point of view, finite spin Drude weight in a non-magnetized sector indicates that certain excitations attain finite dressed magnetization even in the $\mu \to 0$ limit, mirroring the situation in the easy-plane phase of the quantum Heisenberg XXZ spin chain \cite{IN_Drude,superdiffusion_review},
\item from the viewpoint of the Mazur--Suzuki hydrodynamic projection~\cite{Mazur69,Suzuki71}
(for more information see e.g.~\cite{Ilievski12,Doyon2019_projections,Doyon_lectures}) a finite value of $\mathcal{D}$ necessitates the existence of local
(or quasilocal, see Refs.~\cite{Ilievski12,IMP15,quasilocal_review}) conserved quantities with finite spin-charge susceptibility
even at $\mu \to 0$, that is $\langle J\, Q\rangle^{c}_{0} > 0$.
\end{enumerate}

Conserved quantities with a finite overlap with spin current have indeed been found in the gapless phase of the quantum Heisenberg chain
(see Refs.~\cite{Prosen11,PI13,ProsenNPB14,Pereira14} and \cite{quasilocal_review} for a review). They are commonly referred to
as the `$Z$-charges'.
Presently however, the standard local conservation laws $Q^{\pm}_{k;\tau}$ extracted by expanding the trace of
the monodromy matrix as an analytic series (see the derivation in \appref{sec:iso}) all satisfy $\langle J\, Q^{\pm}_{k;\tau}\rangle^{c}_{0} = 0$.
This indeed follows directly from a simple symmetry argument: while all local charges $Q_{k;\tau}$ remain invariant under the spin-reversal transformation, the spin current $J$ flips a sign.

\paragraph*{Diffusion constant.}
The diffusion constant $D$ is finite for all values of $\ani$ provided $\mu > 0$. In the $\mu \to 0$ limit however, it once again diverges
as $D \sim 1/\ani^{\varkappa_{\rm ep}}$ algebraically with an estimated exponent $\varkappa_{\rm ep}=1$, cf. \figref{fig:easy_plane1} (b, inset). We can observe, see \figref{fig:easy_plane20} (a, blue line), a crossover behavior in the vicinity of the isotropic point from ballistic to superdiffusive dynamics. In the easy-plane regime, the stationary profiles (Fig.~\ref{fig:easy_plane21}) have two peaks which are indicative of ballistic spreading. At large $\tau$ (with fixed $\ani$) the two-particle map reduces to
the permutation, see Eq.~\eqref{easy_plane_swap}, meaning that the dynamics becomes non-interacting and hence the structure factor concentrates near the light-cone boundary peaked at
$\ell/t =  \pm 2$.

\section{Conclusion}
\label{sec:conclusion}

We have introduced a new integrable model in discrete space-time, representing classical spins which interact locally via an anisotropic spin-spin interaction. The model can be perceived as an integrable discrete-time analogue of the uniaxailly anisotropic Landau-Lifshitz
field theory. The key ingredient of our algebraic construction is the discrete version of the zero-curvature condition on an auxiliary
light-cone lattice, which implicitly defines the elementary time-propagator. We managed to express the elementary propagator
explicitly as a two-body symplectic transformation by exploiting certain factorization properties of the Lax matrix.
The full propagator takes the form of a classical circuit made of symplectic maps. The two free parameters of the model are the
Trotter time-step and the interaction anisotropy, the latter can be chosen real or purely imaginary
(corresponding to the easy-axis and easy-plane regimes, respectively).


\medskip

Our discrete model provides an efficient explicit integration scheme that manifestly preserves integrability.
As an important immediate physics application, we numerically studied magnetization transport in grand-canonical stationary ensembles at different values
of magnetization density. With only a moderate computational effort, we compute the linear transport coefficients
(the spin Drude weight and spin diffusion constant) with aid of the Kubo formula, both in the easy-axis and easy-plane regimes.

The spin Drude weight is a non-negative quantity. As expected, we found that it only vanishes in (a) the limit of strong polarization
and in (b) the easy-axis regime at zero net magnetization. On sub-ballistic scales, there is a finite diffusive correction characterized
by the spin diffusion constant. According to expectations, we have found that the spin diffusion constant
blows up when approaching the isotropic point (within either regime). This divergence is numerically estimated to be roughly 
inversely proportionally to the anisotropy strength. Close to the isotropic point, we encountered discernible finite-time crossover effects which signal the onset of spin superdiffusion \cite{superdiffusion_review} (characterized by the dynamical exponent $\alpha = 2/3$).

In the easy-plane regime, the spin Drude weight retains a finite value even at zero net magnetization, consistent with the findings of Refs.~\cite{PZ13,DasKPZ}. It is well-known indeed that the same behavior occurs in the the gapless phase of the Heisenberg quantum spin chain,
where the non-vanishing of the spin Drude weight is intimately linked with the existence of the so-called quasilocal conservation laws~\cite{Prosen11,PI13,quasilocal_review} of odd parity under the spin-reversal transformation. 
The most suggestive explanation of this peculiarity is that there exist additional `hidden' (quasi)local conservation laws in the classical spin model as well -- a classical analogue of the $Z$-charges.

\medskip

The hope is that our numerical data can serve as an accurate test of these theoretical predictions.
The spin Drude weight and diffusion constant are in principle amenable to analytic computation, e.g. using the universal formulae for the Drude weights~\cite{DS17,IN_Hubbard} and diffusion constants~\cite{DDB18,Gopalakrishnan18,DDB19} obtained within the formalism of
 generalized hydrodynamics~\cite{BCDF16,CDY16}. This computation however requires the knowledge of the quasiparticle content of the model and the associated thermodynamic state functions \cite{Doyon_lectures}, which are yet to be derived,
e.g. by means of the inverse scattering techiques \cite{Faddeev1987,AblowitzSegur_book, Novikov_book,Babelon_book} adapted to
the fully discrete setting. We leave this task for future research.

\section*{Acknowledgements}

The work has been supported by ERC Advanced grant 694544 -- OMNES and the program P1-0402 of Slovenian Research Agency.  We thank Johannes Schmidt for pointing out a mistake in Eq.~(\ref{reality_condition}).

\appendices
\addtocontents{toc}{\protect\setcounter{tocdepth}{1}}

\section{Semiclassical limit of quantum algebra}
\label{app:semiclassical}

The algebraic structures pertaining to the lattice Landau--Lifshitz model can be derived systematically by taking the
semiclassical limit of a quasi-triangular Hopf algebra (i.e. the `quantum group') $\mathcal{U}_{q}(\mathfrak{su}_{2})$.
For the reader's convenience, we summarize the key steps of the derivation.

\paragraph*{Fundamental commutation relation.}
A natural starting point is to consideer the so-called fundamental commutation (or RLL) relation
of the difference form~\cite{Sklyanin83,FRT88,Korepin_book}
\begin{equation}
R_{12}(\lambda - \lambda')\hat{L}_{13}(\lambda)\hat{L}_{23}(\lambda') = \hat{L}_{23}(\lambda')\hat{L}_{13}(\lambda)R_{12}(\lambda-\lambda'),
\label{RLL}
\end{equation}
representing an operatorial identity on the tensor product of three vector spaces of the form
$\mathcal{V}_{1/2}\otimes \mathcal{V}_{1/2}\otimes \mathcal{V}_{S}$.
We have employed the standard embedding prescription in which the subscript indices specify on which spaces the operators act non-identically.

Lax matrices act on $\mathbb{C}^{2}\otimes \mathcal{V}_{S}$, where $\mathcal{V}_{S}$ pertains to a unitary irreducible representation
of `Sklynanin quantum spins' $\hat{\mathcal{S}}^{a}$ (with $a \in \{0,1,2,3\}$) of dimension $2S+1$. The Lax operator assumes the form
\begin{equation}
L(u) = \mathds{1}\otimes \hat{\mathcal{S}}^{0} + \sum_{a=1}^{3}W_{a}(\lambda)\,\sigma^{a}\otimes \hat{\mathcal{S}}^{a}.
\label{quantum_Lax}
\end{equation}

\paragraph*{Quantum $R$-matrix.}
The $R$-matrix acts on two copies of $\mathcal{V}_{1/2}\cong \mathbb{C}^{2}$ which are regarded as \emph{auxiliary} fundamental spins (subsequently represented by Pauli matrices $\sigma^{a}$). The $R$-matrix swaps (intertwines) the spectral parameters of the Lax operators.\\
The $R$-matrix has the $6$-vertex form~\cite{Korepin_book}
\begin{equation}
R(\lambda) = \mathds{1} + \sum_{a=1}^{3}W_{a}(\lambda)\,\sigma^{a}\otimes \sigma^{a},
\label{R6vertex}
\end{equation}
with trigonometric amplitudes (Boltzmann weights)
\begin{equation}
W_{1}(\lambda) = W_{2}(\lambda) = \frac{\sinh{(\eta)}}{\sinh{(\lambda + \eta)}},\qquad
W_{3}(\lambda) = \frac{\tanh{(\eta)}}{\tanh{(\lambda + \eta)}},
\end{equation}
where
\begin{equation}
q=e^{\eta} \in \mathbb{R}_{+},
\end{equation}
is the quantum deformation parameter. The $6$-vertex quantum $R$-matrix, cf. Eq.~\eqref{R6vertex}, is a specialization of
more general $8$-vertex elliptic $R$-matrix introduced by Sklyanin \cite{Sklyanin89XYZ}, obtained by sending the elliptic modulus to zero.

When embedded in $\mathcal{V}_{1/2}\otimes \mathcal{V}_{1/2}\otimes \mathcal{V}_{1/2}$, the $R$-matrix satisfies the quantum
Yang--Baxter relation~\cite{Baxter_book,Korepin_book}
\begin{equation}
R_{12}(\lambda - \lambda')R_{13}(\lambda)R_{23}(\lambda') = R_{23}(\lambda')R_{13}(\lambda)R_{12}(\lambda-\lambda').
\label{RRR}
\end{equation}

\paragraph*{Quantum Sklyanin algebra.}
The fundamental commutation relation \eqref{RLL} is equivalent to the Sklyanin quadratic algebra
\begin{equation}
[\hat{\mathcal{S}}^{0},\hat{\mathcal{S}}^{a}] = \ii \, \mathbb{J}_{bc}[\hat{\mathcal{S}}^{b},\hat{\mathcal{S}}^{c}]_{+},\qquad
[\hat{\mathcal{S}}^{a},\hat{\mathcal{S}}^{b}] = \ii [\hat{\mathcal{S}}^{0},\hat{\mathcal{S}}^{c}]_{+},
\end{equation}
which, in the general elliptic case, involves four generators $\hat{\mathcal{S}}^{a}$, with $a\in \{1,2,3,4\}$, and three `structure constants'
$\mathbb{J}_{ab}\equiv -(\mathbb{J}_{a}-\mathbb{J}_{b})/\mathbb{J}_{c}$ (with indices $a,b,c$ mutually distinct) obeying the constraint
$\mathbb{J}_{12}+\mathbb{J}_{23}+\mathbb{J}_{31}+\mathbb{J}_{12}\mathbb{J}_{23}\mathbb{J}_{31}=0$.
The Boltzmann weights specify an algebraic curve $W^{2}_{a} - W^{2}_{b}=\mathbb{J}_{ab}(W^{2}_{c}-1)$.
The elliptic Sklyanin algebra admits two independent quadratic Casimir operators
\begin{equation}
\hat{\mathscr{C}}_{0} = \sum_{a=0}^{3}\big(\hat{\mathcal{S}}^{a}\big)^{2},\qquad
\hat{\mathscr{C}}_{1} = \sum_{a=1}^{3}\mathbb{J}_{a}\big(\hat{\mathcal{S}}^{a}\big)^{2}.
\end{equation}

We are interested specifically in the trigonometric limit $\mathbb{J}_{12} \to 0$, where the Sklyanin algebra reduces to the
quantum algebra $\mathcal{U}_{q}(\mathfrak{su}_{2})$, representing a $q$-deformed quantum spin.
In this limit, one finds $\hat{\mathscr{C}}_{0}-\hat{\mathscr{C}}_{1}=1$, which leaves us with a single non-trivial Casimir operator
of the form
\begin{equation}
\hat{\mathcal{C}}_{q} \equiv q^{-1}\hat{\mathcal{K}}^{2} + q\hat{\mathcal{K}}^{-2} + (q-q^{-1})\hat{\mathcal{S}}^{+}\hat{\mathcal{S}}^{-}.
\end{equation}
Moreover, writing $\kappa \equiv \sqrt{\mathbb{J}_{23}}=\sqrt{(\mathbb{J}_{3}-\mathbb{J}_{2})/\mathbb{J}_{1}}$ and introducing
\begin{equation}
\hat{\mathcal{S}}^{\pm}\equiv \hat{\mathcal{S}}^{1}\pm \ii \hat{\mathcal{S}}^{2},\qquad
\hat{\mathcal{K}} \equiv \hat{\mathcal{S}}^{0} + \kappa\,\hat{\mathcal{S}}^{3} = q^{\hat{\mathcal{S}}^{3}},
\end{equation}
the defining commutation relations of the trigonometric Sklyanin algebra read
\begin{equation}
[\hat{\mathcal{K}},\hat{\mathcal{S}}^{\pm}] = \pm \kappa [\hat{\mathcal{K}},\hat{\mathcal{S}}^{\pm}]_{+},\qquad
[\hat{\mathcal{S}}^{+},\hat{\mathcal{S}}^{-}] = \frac{1}{\kappa}\big(\hat{\mathcal{K}}^{2} - \hat{\mathcal{K}}^{-2}\big).
\label{quantum_Sklyanin_algebra_trig}
\end{equation}
One can readily verify that these are indeed equivalent to standard $\mathcal{U}_{q}(\mathfrak{su}_{2})$ algebraic relations
upon introducing the $q$-deformation parameter $q=(1+\kappa)/(1-\kappa)$ and rescaling the generators,
\begin{equation}
\hat{\mathscr{S}}^{\pm} \equiv \tfrac{1}{2}\sqrt{(1+\kappa)(1-\kappa)}\,\hat{\mathcal{S}}^{\pm},\qquad \hat{\mathscr{K}} \equiv \hat{\mathcal{K}},
\end{equation}
yielding
\begin{equation}
\hat{\mathscr{K}}\,\hat{\mathscr{S}}^{\pm} = q^{\pm 1}\,\hat{\mathscr{S}}^{\pm}\,\hat{\mathscr{K}},\qquad
[\hat{\mathscr{S}}^{+},\hat{\mathscr{S}}^{-}] = \frac{\hat{\mathscr{K}}^{2}-\hat{\mathscr{K}}^{-2}}{q-q^{-1}}.
\label{Uqsl2_standard}
\end{equation}

\subsection{Semiclassical limit}

We now perform the semiclassical limit. This is achieved, at the algebraic level, by expanding the fundamental commutation relation about the `classical point' $q=1$. Following Refs.~\cite{Sklyanin83,Sklyanin88}, we parameterize $q=e^{\eta}$, put $\eta = \hbar\,\varrho$ and
subsequently expand the quantum $R$-matrix to the leading order in $\hbar$,
\begin{equation}
R(\lambda) = \mathds{1} - 2 \ii \hbar \, r(\lambda) + \mathcal{O}(\hbar^{2}),
\end{equation}
where~\footnote{There is freedom to choose the overall scale of the $r$-matrix. Here we adopt the common convention \cite{Faddeev1987},
which amounts to fixing a particular normalization of the Sklyanin Poisson algebra, cf. Eq.~\eqref{classical_Sklyanin}.
Notice that our normalization differs from that of Ref.~\cite{Sklyanin83}.}
\begin{equation}
r(\lambda) = -\frac{1}{2}\sum_{a=1}^{3}w_{a}(\lambda)\,\sigma^{a}\otimes \sigma^{a},
\label{classical_r}
\end{equation}
is known as the classical $r$-matrix~\cite{Sklyanin83,Drinfeld88,Sklyanin88}.
The classical amplitudes (Boltzmann weights) $w_{a}(\lambda)$ are similarly found by expanding the $W$-functions
$W_{a}(\lambda)=\ii \hbar\,w_{a}(\lambda)+\mathcal{O}(\hbar^{2})$, and structure constatnts
$\mathbb{J}_{a}=1-\hbar J_{a}+\mathcal{O}(\hbar^{2})$, yielding
\begin{equation}
w_{1,2}(\lambda) = \frac{\varrho}{\sin{(\lambda)}},\qquad
w_{3}(\lambda) = \frac{\varrho}{\tan{(\lambda)}},
\end{equation}
with anisotropy parameter
\begin{equation}
\varrho \equiv \sqrt{J_{3}-J_{1}} \in \mathbb{R}_{+}
\end{equation}
The classical structure constants now obey $J_{12}+J_{23}+J_{31}=0$,
and prescribe an algebraic curve $w^{2}_{a}(\lambda) - w^{2}_{b}(\lambda)=J_{b}-J_{a}$.

The classical $r$-matrix satisfies the classical analogue of the Yang--Baxter relation~\cite{Sklyanin83,Drinfeld88}
\begin{equation}
[r_{12}(\lambda-\lambda'),r_{13}(\lambda)] + [r_{12}(\lambda-\lambda'),r_{23}(\lambda')] + [r_{13}(\lambda),r_{23}(\lambda')] = 0,
\label{cYBE}
\end{equation}
which can be readily deduced by expanding the Yang--Baxter relation \eqref{RRR} to the leading order $\mathcal{O}(\hbar^{2})$.

\paragraph*{Sklyanin quadratic Poisson algebra.}

In order to deduce the semiclassical limit of the fundamental commutation relation, one has to send $\hbar \to 0$ while keeping
$\varrho$ and $\hbar\,S$ constant, thereby fixing the size of classical variables. This is achieved as follows.
By expanding Eq.~\eqref{RLL} to the leading order $\mathcal{O}(\hbar)$ and subsequently replacing commutators with Poisson brackets
by virtue of the canonical correspondence principle~\footnote{Normalization of the Poisson bracket is set by the choice of
the classical $r$-matrix, cf. Eq.~\eqref{classical_r}.}
\begin{equation}
[\circ,\circ]\to 2\ii \hbar \{\circ,\circ\},
\label{correspondence_principle}
\end{equation}
the fundamental commutation relation reduces to the Sklyanin quadratic bracket~\cite{Sklyanin83}
\begin{equation}
\big\{L(\lambda)\stackrel{\otimes}{,}L(\lambda')\big\} = \big[r(\lambda-\lambda'),L(\lambda)\otimes L(\lambda')\big],
\label{Sklyanin_bracket}
\end{equation}
with the classical Lax matrix~\cite{Sklyanin83}
\begin{equation}
L(\lambda) = \mathds{1}\,\mathcal{S}^{0} + \frac{1}{\ii}\sum_{a}w_{a}(\lambda)\sigma^{a}\mathcal{S}^{a},
\end{equation}
where we have simultaneously introduced classical `Sklyanin variables'~\footnote{Our normalization here slightly differs
from the one in Ref.~\cite{Faddeev1987} in that $\mathcal{S}^{a}$ for $a\in \{1,2,3\}$ were rescaled by a factor of $\varrho^{-1}$.}
\begin{equation}
\hat{\mathcal{S}}^{0} \to \hbar^{-1}\mathcal{S}^{0},\qquad
\hat{\mathcal{S}}^{a} \rightarrow \hbar^{-1}\varrho^{-1}\mathcal{S}^{a},\quad {\rm for} \quad a\in \{1,2,3\}.
\end{equation}
Sklyanin quadratic bracket \eqref{Sklyanin_bracket} provides a compact algebraic representation
for the following \emph{quadratic} Poisson algebra~\cite{Sklyanin83}
\begin{equation}
\{\mathcal{S}^{0},\mathcal{S}^{3}\} = 0,\qquad
\{\mathcal{S}^{0},\mathcal{S}^{a}\} = \varrho^{-1}\epsilon_{abc} \, J_{bc} \, \mathcal{S}^{b}\mathcal{S}^{c},\qquad
\{\mathcal{S}^{a},\mathcal{S}^{b}\} = -\varrho \, \epsilon_{abc} \, \mathcal{S}^{0}\mathcal{S}^{c},
\label{classical_Sklyanin}
\end{equation}
with classical structure constants $J_{ab} \equiv J_{a}-J_{b}$. Poisson algebra \eqref{classical_Sklyanin} admits
two independent Casimir functions
\begin{equation}
\mathcal{C}_{0} \equiv \sum_{a=1}^{3}\big(\mathcal{S}^{a}\big)^{2},\qquad
\mathcal{C}_{1} \equiv \big(\mathcal{S}^{0}\big)^{2} - \big(\mathcal{S}^{3}\big)^{2}.
\label{classical_Casimirs}
\end{equation}
By prescribing their values $c_{0}$ and $c_{1}$, respectively, variables $\mathcal{S}^{a}$ takes values
on a two-dimensional non-degenerate Poisson submanifold. By fixing them to
\begin{equation}
c_{0} = \sinh^{2}(\varrho),\qquad
c_{1} = 1,
\end{equation}
the manifold is topologically equivalent to a two-sphere; $\{\mathcal{S}^{a}\}_{a=1}^{3}$ lie on a two-sphere,
whereas $\mathcal{S}^{0} > 0$ is uniquely determined by $\mathcal{S}^{3}$ through $\mathcal{C}_{1}$.

Introducing linear combinations $\mathcal{K}^{\pm}\equiv \mathcal{S}^{0} \pm \mathcal{S}^{3}$
and putting $\mathcal{K}\equiv \mathcal{K}^{+}$, we have $\mathcal{K}^{+}\mathcal{K}^{-}=\mathcal{C}_{1}$, yielding~\cite{Sklyanin89XXZ}
\begin{equation}
\mathcal{A}_{q}:\qquad
\{\mathcal{K},\mathcal{S}^{\pm}\} = \mp \ii \varrho\,\mathcal{S}^{\pm}\mathcal{K},\qquad
\{\mathcal{S}^{+},\mathcal{S}^{-}\} = -\frac{\ii \varrho}{2}(\mathcal{K}^{2}-\mathcal{K}^{-2}).
\end{equation}
The above Poisson relation also follow from quantum algebraic relations \eqref{quantum_Sklyanin_algebra_trig}
upon rescaling $\kappa$ as $\kappa = \hbar\,\varrho$, applying the canonical corespondence \eqref{correspondence_principle},
and finally sending $\hbar \to 0$.  Poisson algebra $\mathcal{A}_{q}$, with a pair of Casimir functions $\mathcal{C}_{0}$, $\mathcal{C}_{1}$, can be understood as the classical counterpart of $\mathcal{U}_{q}(\mathfrak{su}_{2})$.

\section{Factorization in Weyl variables}
\label{app:weyl_propagator}

We provide explicit prescriptions for the elementary exchanges of Weyl matrix $U$ and $V$ that appear in the decomposition of the Lax matrix (cf. \secref{sec:discrete_model}), written in the form of symplectic transformations on Weyl pairs ($x_\ell,y_\ell$):\\

\begin{enumerate}[label=(\roman*)]

\item $U_1-U_2$ exchange \eqref{U_swap}
\begin{align}
x'_1 &= x_1, \label{swapU1} \\
x'_2 & = x_2  + (u_1 - u_2) \frac{x_1x_2 v_2(u_1 + u_2 y_2^2) - x_2^2(1+ u_1 u_2 v_2^2 y_2^2)}{x_1 u_1 u_2 v_2  (1-y_2^2) + u_2 x_2 (u_1^2 v_2^2 y_2^2 - 1)} ,\\
 (y'_1)^2 & = y_1^2 - (u_1^2 - u_2^2) \frac{v_2 x_1 y_1^2  (y_2^2 -1)}{u_1 x_2(u_2^2 v_2^2 y_2^2 - 1) + x_1 v_2 u_2^2(1 - y_2^2 )},\\
 (y'_2)^2 & = y_2^2 + (u_1^2 - u_2^2) \frac{v_2 x_1 y_2^2  (y_2^2 -1)}{u_1 x_2(u_2^2 v_2^2 y_2^2 - 1) + x_1 v_2 u_1^2(1 - y_2^2 )},
 \label{swapU4}
\end{align}
and in `Trotter parametrization' \eqref{uv_Trotter_param} \footnote{Where here and in the following we use $w = e^{\ii \varrho \tau}$.}
\begin{align}
{x'}_1 &= x_1, \\
{x'}_2 & = x_2 \frac{x_1 (y_2 w^{-1}  - wy_2^{-1}) + x_2(\nu^{-2}y_2^{-1} - y_2\nu^2)}{x_1(y_2-y_2^{-1}) + x_2(w^{-1}\nu^{-2}y_2^{-1} - y_2 w \nu^2)} ,\\
 (y'_1)^2 &=  y_1^2 \frac{x_1(y_2-y_2^{-1})w + x_2(y_2^{-1}\nu^{-2} - y_2 \nu^2)}{x_1(y_2-y_2^{-1})w^{-1} + x_2(y_2^{-1}\nu^{-2} - y_2 \nu^2)},\\
 (y'_2)^2 &= y_2^2 \frac{x_1(y_2-y_2^{-1})w^{-1} + x_2(y_2^{-1}\nu^{-2} - y_2 \nu^2)}{x_1(y_2-y_2^{-1})w + x_2(y_2^{-1}\nu^{-2} - y_2 \nu^2)},
\end{align}

\item $V_1 - V_2$ exchange \eqref{V_swap}

\begin{align}
x'_1 & = x_1  - (v_1 - v_2) \frac{x_1x_2 u_1(v_2 + v_1 y_1^2) - x_1^2(1+ v_1 v_2 u_1^2 y_1^2)}{x_2 v_1 v_2 u_1  (1-y_1^2) + v_1 x_1 (u_1^2 v_2^2 y_1^2 - 1)}, \label{swapV1}\\
x'_2 &= x_2, \\
(y'_1)^2 & = y_1^2 - (v_1^2 - v_2^2) \frac{u_1 x_2 y_1^2  (y_1^2 -1)}{v_2 x_1(u_1^2 v_1^2 y_1^2 - 1) + x_2 u_1 v_2^2(1- y_1^2 )},\\
(y'_2)^2 & = y_2^2 + (v_1^2 - v_2^2) \frac{u_1 x_2 y_2^2  (y_1^2 -1)}{v_2 x_1(u_1^2 v_1^2 y_1^2 - 1) + x_2 u_1 v_1^2(1 - y_1^2 )},
\label{swapV4}
\end{align}
and in `Trotter parametrization' \eqref{uv_Trotter_param}
\begin{align}
{x}'_1 &= x_1\frac{x_1(y_1^{-1}\nu^{-2} - y_1\nu^2) + x_2  (y_1w^{-1}  - wy_1^{-1}) }{x_1  (y_1^{-1}w^{-1}\nu^{-2} - y_1w \nu^2) + x_2 (y_1-y_1^{-1})},\\
{x}'_2 &= x_2, \\
 (y'_1)^2 &=  y_1^2 \frac{x_1(y_1^{-1}\nu^{-2} - \nu^2 y_1) + x_2 w^{-1} (y_1 - y_1^{-1})}{x_1(y_1^{-1}\nu^{-2} - \nu^2 y_1) + x_2 w (y_1 - y_1^{-1})},\\
 (y'_2)^2 &= y_2^2 \frac{x_1(y_1^{-1}\nu^{-2} - \nu^2 y_1) + x_2 w (y_1 - y_1^{-1})}{x_1(y_1^{-1}\nu^{-2} - \nu^2 y_1) + x_2 w^{-1} (y_1 - y_1^{-1})}.
\end{align}

\end{enumerate}
The combined $U-V$ exchange (in Trotter parametrization \eqref{uv_Trotter_param}) gives the two-body map $\Phi$
\begin{align}
x_1' &= x_1\frac{x_1 Y_1(\nu, 0) - x_2  Y_1(1,- \tau) }{x_1 Y_1(\nu,  \tau)  - x_2 Y_1(1, 0)}, &\qquad
x_2' &=  x_2\frac{x_1 Y_2(1, - \tau) - x_2 Y_2(\nu, 0)}{x_1Y_2(1, 0)- x_2Y_2(\nu,  \tau)},\\
y_1'^2 &= y_1^2 \frac{f^-_2}{f_1^+},&\qquad
 y_2'^2 &= y_2^2\frac{f_1^+}{f_2^-},
\label{xy_eom}
\end{align}
with
\begin{align}
f_j^\pm &= x_{j\pm1}^2 wY_j(1, 0)Y_{j\pm1}(\nu, 0) + x_j^2 w^{-1}Y_{j\pm1}(1, 0)Y_j(\nu, 0) \nonumber \\
&- x_{j\pm1} x_j\left( Y_{j\pm1}(\nu, 0) Y_j(\nu, 0) +Y_{j\pm1}(1, 0) Y_j(1, 0) - y_j y_{j\pm 1}^{-1} (w-w^{-1})^2 \right),\\
Y_j(\nu, \tau) &= \nu^2 y_j  w -  \nu^{-2} y_j^{-1} w^{-1}.
\end{align}

\section{Isospectral flows and local conservation laws}
\label{sec:iso}
\noindent 

A key implication of the discrete zero-curvature condition \eqref{eqn:ZC_abstract} is the existence of infinitely many conserved quantities
which are in mutual involution. As we shortly demonstrate, these are a corollary of isospectrality of the transfer function.
We accordingly introduce a staggared monodromy matrix $T_{\tau}(\lambda)\equiv T_\tau(\{\vec{S}_{\ell}\};\lambda)$, defined
as a right-to-left ordered product of Lax matrices along a zig-zag path (i.e. initial sawtooth) of length $L$
\begin{equation}
T_\tau(\lambda) =  L \big(\vec{S}_{L};\lambda^{+}\big) L \big(\vec{S}_{L-1};\lambda^{-}\big)\cdots  L \big(\vec{S}_{2};\lambda^{+}\big) L \big(\vec{S}_{1};\lambda^{-}\big).
\label{monodromy}
\end{equation}
where $\lambda^{\pm}\equiv \lambda \pm \tau/2$. Here and subsequently we leave the dependence on anisotropy $\varrho$ implicit.
By tracing out the auxiliary $\mathbb{C}^{2}$ space, we obtain the transfer function
\begin{equation}
t_{\tau}(\lambda) = \textrm{Tr}\, T_\tau(\lambda).
\end{equation}

\paragraph*{Isospectral evolution.}
By virtue of the zero-curvature condition \eqref{eqn:ZC_abstract} and cyclic property of the trace, it follows that
\begin{equation}
t_\tau(\lambda)\circ \Phi^{\rm even}_{\tau} = t_{-\tau}(\lambda), \qquad
t_{-\tau}(\lambda)\circ \Phi^{\rm odd}_{\tau} = t_{\tau}(\lambda),
\end{equation}
which in combination imply invariance under the time propagation \eqref{full_propagator}
\begin{equation}
t_\tau(\lambda) \circ \Phi^{\rm full}_{\tau} =  t_\tau(\lambda).
\label{trace_invariance}
\end{equation}
The quadratic bracket \eqref{Sklyanin_bracket} can be immediately lifted onto the level of monodromies
\begin{equation}
\big\{T_\tau(\lambda) \stackrel{\otimes}{,} T_\tau(\mu) \big\} = \left [r(\lambda - \mu), T_\tau(\lambda) \otimes T_\tau(\mu)  \right ],
\end{equation}
By subsequently taking the trace over $\mathcal{C}^{2}$, one readily finds that transfer functions Poisson commute,
\begin{equation}
\big\{ t_\tau(\lambda), t_\tau(\lambda') \big\} = 0,
\label{trace_involution}
\end{equation}
for any pair of complex spectral parameters $\lambda,\lambda'\in \mathbb{C}$ and fixed $\tau$. Preservation of the
characteristic polynomial of $T_{\tau}(\lambda)$ is referred to as isospectrality.

\paragraph*{Local conservation laws.}
Because the transfer function $t_{\tau}(\lambda)$ depends analytically on $\lambda$, it can be used as a generating function
of integrals of motion. In the thermodynamic limit, $L\to \infty$, this provides infinitely many functionally independent conserved quantities.
Expanding $t_\tau(\lambda)$ in a power series in $\lambda$ nonetheless produces complex-valued and \emph{non-local} conserved quantities.
In order to construct real conservation laws, it is useful to utilize the following involutions of the Lax matrix
\begin{equation}
\overline{L(\lambda)} = \sigma^{2} \, L(\overline{\lambda}) \, \sigma^{2}, \qquad
L^{T}(\lambda) = \sigma^{2} \, L(-\lambda) \, \sigma^{2}, 
\end{equation}
which likewise apply to the monodromy matrix $T_{\tau}(\lambda)$. It follows that the moduli of the traces are also in mutual involution,
and their expansion in $\lambda$ thus yields manifestly real conserved quantities.

One can in fact extract two seperate infinite towers of \emph{local} conserved quantities, denoting them by $Q^{\pm}_{k;\tau}$, using the standard approach (see, for example, Ref.~\cite{Faddeev1987}) by expanding the logarithmic derivatives of the modulus of $t_{\tau}(\lambda)$ around either of the two distinguished points $\lambda_{0}^{\pm}= \ii \mp \frac{\tau}{2}$ at which the Lax matrix degenerates (becoming a projector of rank one),
\begin{equation}
{\rm Det}\,L\Big(\lambda_{0}^{\pm}  \Big)  = 0,\qquad
L\Big(\lambda_{0}^{\pm} \Big) = \ket{\alpha_{\pm}}\bra{\beta_\pm},
\end{equation}
yielding two independent sequences of local charges
\begin{align} 
Q^{\pm}_{k;\tau} = \frac{\partial^k}{\partial \lambda^k} \log |t_\tau(\lambda)|^2 \Bigg|_{\lambda = \lambda_0^\pm}.
\end{align}
Taking the continuous time limit $\tau \rightarrow 0$, both branches merge into the local conseration laws
of the anisotropic lattice Landau-Lifshitz model.

\paragraph*{Parity symmetry.}

The transfer function $t_{\tau}(\lambda)$ is $\mathbb{Z}_{2}$-invariant under the spin-reversal transformation
($\pi$-rotation of the canonical spin ${\bf S}$ around the $x$-axis)
\begin{equation}
\mathcal{F}_{1}:\qquad S^{3} \mapsto -S^{3},\qquad S^{\pm} \mapsto S^{\mp},
\end{equation} 
implying $\mathcal{F}_{1}(\mathcal{K})=\mathcal{K}^{-1}$. At the level of the Lax matrix, $\mathcal{F}_{1}$ represents
conjugation by Pauli matrix $\sigma^{1}$,
\begin{equation}
\mathcal{F}_{1}:\qquad L(\lambda) \mapsto \sigma^{1} \, L(\lambda) \, \sigma^{1},\qquad T(\lambda) \mapsto \sigma^{1} \, T(\lambda) \, \sigma^{1},
\end{equation}
which implies invariance $\mathcal{F}_{1}(t_{\tau}(\lambda))=t_{\tau}(\lambda)$ and hence charges $Q^{\pm}_{k;\tau}$ are all
invariant under $\mathcal{F}_{1}$.

\section{Reductions}
\label{app:reductions}

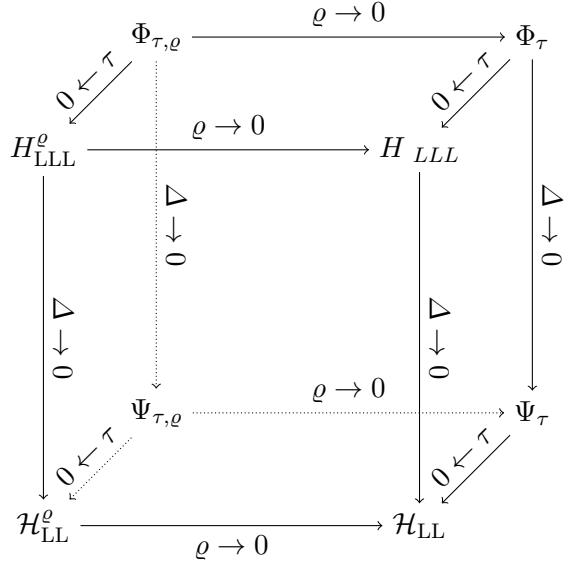
\begin{figure}[h!]
\centering
\tikzset{node distance=5cm, auto}
\begin{tikzpicture}[%
  back line/.style={densely dotted},
  cross line/.style={preaction={draw=white, -,line width=6pt}}]
  \node (A) {$H_{\rm LLL}^{\varrho}$};
  \node [right of=A] (B) {$H_{\ LLL}$};
  \node [below of=A] (C) {$\mathcal{H}_{\rm LL}^{\varrho}$};
  \node [right of=C] (D) {$\mathcal{H}_{\rm LL}$};

  \node (A1) [right of=A, above of=A, node distance=1.5cm]{$\Phi_{\tau,\varrho}$};
  \node [right of=A1] (B1) {$\Phi_{\tau}$};
  \node [below of=A1] (C1) {$\Psi_{\tau,\varrho}$};
  \node [right of=C1] (D1) {$\Psi_{\tau}$};

  \draw[->] (A1) to node {$\varrho \rightarrow 0$} (B1);
  \draw[->] (A1) to node [rotate=45,anchor=south] {$0 \leftarrow \tau$} (A);
  \draw[->] (B1) to node [rotate=45,anchor=south] {$0 \leftarrow \tau$} (B);
  \draw[->] (A) to node  {$\varrho \rightarrow 0$} (B);

  \draw[->, back line] (C1) to node {$\varrho \rightarrow 0$} (D1);
  \draw[<-, back line] (C) to node [rotate=45,anchor=south] {$0 \leftarrow \tau$} (C1);
  \draw[<-] (D) to node [rotate=45,anchor=south]  {$0 \leftarrow \tau$} (D1);
  \draw[->] (C) to node [swap]  {$\varrho \rightarrow 0$} (D);

  \draw[->, back line] (A1) to node [rotate=-90,anchor=south] {$\varDelta \rightarrow 0$} (C1);
  \draw[->] (A) to node [rotate=-90,anchor=south] {$\varDelta \rightarrow 0$} (C);
  \draw[->] (B) to node [rotate=-90,anchor=south] {$\varDelta \rightarrow 0$} (D);
  \draw[->] (B1) to node [rotate=-90,anchor=south]  {$\varDelta \rightarrow 0$} (D1);
\end{tikzpicture}
\caption{Reductions of the anisotropic discrete-time propagator $\Phi_{\tau,\varrho}$ (upper-left corner),
obtained via taking various combinations of $ \varrho, \varDelta, \tau \rightarrow 0$ limits. The complete reduction obtained by
taking all three limit yields the isotropic Landau-Lifshitz field theory $\mathcal{H}_{\rm LL}$ (lower-right corner).}
\label{fig:limits}
\end{figure}

The anisotropic model of interacting spin in discrete space-time, introduced in \secref{sec:discrete_model}, admits several important
reductions. There are three distinct parameters that can be taken to zero: anisotropy $\varrho$, discrete time-step $\tau$,
and lattice spacing $\varDelta$ (which we have kept fixed $\varDelta=1$ throughout the text).
The full set of models which descend from $\Phi_{\tau,\varrho}$ can be therefore arranged on the vertices of a cube,
as depicted in \figref{fig:limits}. We briefly describe them below.~\footnote{We shall omit discussing
$\Psi_\tau^{\varrho}$ and $\Psi_\tau$, which correspond to (anisotropic) `kicked' field theories in discrete time.}

\paragraph*{$\Phi_\tau$ -- isotropic discrete space-time model.}

By letting the anisotropy parameter to zero, $\varrho \rightarrow 0$, the anisotropy Lax matrix given by Eq.~\eqref{Lax1} becomes
the Lax matrix of the isotropic lattice Landau--Lifshitz model~\cite{Ishimori82,Sklyanin83}
\begin{equation}
L^{(0)}(\lambda, \vec{S}) = \mathds{1} + \frac{1}{\ii \lambda} \vec{S} \cdot \boldsymbol{\sigma},
\end{equation}
where $\boldsymbol{\sigma} = (\sigma^1, \sigma^2, \sigma^3)$ is a vector of Pauli matrices.
The corresponding discrete zero-curvature condition around a plaquette,
\begin{equation}
L^{(0)}_2(\lambda + \tau/2)L^{(0)}_1(\lambda - \tau/2)
= L^{\prime (0)}_2(\lambda-\tau/2)L^{\prime (0)}_1(\lambda + \tau/2),
\label{isotropic_ZC}
\end{equation}
generates a two-body symplectic propagator \cite{krajnik2020kardar}
\begin{align}
\Phi_{\tau}  (\vec{S}_1, \vec{S}_2) &=  \frac{1}{s^2 + \tau^2} \left(s^2\mathbf{S}_1 + \tau^2\vec{S}_2 + \tau \vec{S}_1\times\vec{S}_2, s^2\vec{S}_2 + \tau^2\vec{S}_1 + \tau \vec{S}_2\times\vec{S}_1 \right),\\
s^2 &= \frac{1}{2} \left(1 + \vec{S}_1 \cdot \vec{S}_2 \right).
\label{discrete_isotropic_eom}
\end{align}

\paragraph{$H_{\rm LLL}^{\varrho}$ -- uniaxially anisotropic lattice Landau-Lifshitz model.}
By sending $\tau \rightarrow 0$, the discrete-time propagator becomes the equation of motion in continuous time,
\begin{equation}
\frac{\dd}{\dd t} \mathcal{S}_{\ell}^{a} = \left \{ \mathcal{S}^{a}, H_{\rm LLL}^{\varrho} \right \},
\label{LLL_eom}
\end{equation}
generated by a Hamiltonian
\begin{equation}
H_{\rm LLL}^{\varrho} \simeq \sum_{\ell=1}^{L} \log \left[ \sinh^2 \varrho\, \mathcal{S}^{0}_{\ell}\mathcal{S}^{0}_{\ell+1} + \mathcal{S}^{1}_{\ell}\mathcal{S}^{1}_{\ell+1} + \mathcal{S}^{2}_{\ell} \mathcal{S}^{2}_{\ell+1} + \cosh^2 \varrho \mathcal{S}^{3}_{\ell}\mathcal{S}^{3}_{\ell+1}\right],
\end{equation}

\paragraph{$H_{\rm LLL}$ -- isotropic lattice Landau-Lifshitz model.} 

Starting either from Eq.~\eqref{discrete_isotropic_eom} and sending $\tau \rightarrow 0$, or from Eq.~\eqref{LLL_eom}
and sending $\varrho \rightarrow 0$, one ends up with the Hamiltonian dynamics of the isotropic lattice Landau-Lifshitz model.
In terms of canonical spins ${\bf S}$, the equations of motion read~\cite{Ishimori82}
\begin{equation}
\frac{\dd }{\dd t}\vec{S}_\ell = \left\{ \vec{S}_\ell, H_{\textrm{LLL}} \right \} = 
\frac{\vec{S}_{\ell}\times\vec{S}_{\ell-1}}{1+\vec{S}_{\ell}\cdot \vec{S}_{\ell-1}}
+\frac{\vec{S}_{\ell}\times\vec{S}_{\ell+1}}{1+\vec{S}_{\ell}\cdot \vec{S}_{\ell+1}},
\end{equation}
with a Hamiltonian of the form $H_{\textrm{LLL}} = \sum_{\ell=1}^{L} \log \left(1 + \vec{S}_\ell \cdot \vec{S}_{\ell+1} \right)$.

\paragraph{$\mathcal{H}^{\varrho}_{\rm LL}$ -- anisotropic Landau--Lifshitz field theory.}

To obtain the full continuum limit, we reinstate the lattice spacing $\varDelta$ and subsequently retain smooth lattice spin configurations
$\vec{S}_\ell = \vec{S}(x=\Delta \ell)$ by keeping only variations at the leading non-trivial order $\mathcal{O}(\varDelta^{2})$:
\begin{equation}
\vec{S}_{\ell \pm 1} = \vec{S}(x) \pm \varDelta \partial_{x} \vec{S} + \frac{1}{2}\varDelta^{2} \, \partial_{x}^{2} \vec{S} + \mathcal{O}(\varDelta^{3}).
\end{equation}
Upon taking the continuum limit $\varDelta \to 0$, both the Sklyanin variables and the anisotropy parameter 
have to be accordingly rescaled by the lattice spacing, namely
\begin{equation}
\mathcal{S}^{0} = 1 + \mathcal{O}(\varDelta^{2}), \qquad
\mathcal{S}^j = \varDelta \, \vec{S}^{a} + \mathcal{O}(\varDelta^{2}) \quad {\rm for}\quad a=1,2,3,
\end{equation}
and $\varrho \rightarrow \varDelta \, \varrho$. This way, one recovers the anisotropic Landau-Lifshitz field theory~\cite{Sklyanin79,Faddeev1987}
\begin{equation}
\mathcal{H}^\varrho_{\textrm{LL}} =  \frac{1}{2}\int_{0}^{L} \dd x \, \left[\left(\partial_{x}\vec{S}\right)^2 -  \vec{S}\cdot \mathcal{J} \vec{S} \right], \qquad \mathcal{J} = {\rm diag}(J_{1},J_{2},J_{3}),
\label{anisotropic_field_theory}
\end{equation}
(where $J_{1}=J_{2}$ and $J_{3}-J_{1}= \varrho^{2}$) with the equation of motion~\cite{Sklyanin79,Faddeev1987}
\begin{equation}
\partial_{t} \vec{S}(x) = \{\vec{S}(x), \mathcal{H}^\varrho_{\textrm{LL}} \}
= \vec{S} \times \partial_x^2 \vec{S} + \vec{S} \times \mathcal{J} \vec{S}.
\end{equation}

\paragraph{$\mathcal{H}_{\rm LL}$ -- isotropic Landau-Lifshitz field theory.}

By finally sending $\varrho \rightarrow 0$,
Eq.~\eqref{anisotropic_field_theory} becomes the isotropic Landau-Lifshitz field theory~\cite{Takhtajan77,Faddeev1987}
\begin{equation}
\mathcal{H}_{\rm LL} = \frac{1}{2}\int_{0}^{L} \dd x \,\left(\partial_x\vec{S}\right)^2, \qquad
\partial_{t} \vec{S} = \{\vec{S}, \mathcal{H}_{\rm LL} \} = \vec{S} \times \partial_x^2 \vec{S}.
\end{equation}\\

\bibliography{LLLMM}

\begin{thebibliography}{100}
\providecommand{\url}[1]{\texttt{#1}}
\providecommand{\urlprefix}{URL }
\expandafter\ifx\csname urlstyle\endcsname\relax
  \providecommand{\doi}[1]{doi:\discretionary{}{}{}#1}\else
  \providecommand{\doi}{doi:\discretionary{}{}{}\begingroup
  \urlstyle{rm}\Url}\fi
\providecommand{\eprint}[2][]{\url{#2}}

\bibitem{PhysRevLett.91.147902}
G.~Vidal,
\newblock \emph{Efficient classical simulation of slightly entangled quantum
  computations},
\newblock Phys. Rev. Lett. \textbf{91}, 147902 (2003),
\newblock \doi{10.1103/PhysRevLett.91.147902}.

\bibitem{PhysRevLett.93.207204}
F.~Verstraete, J.~J. Garc\'{\i}a-Ripoll and J.~I. Cirac,
\newblock \emph{Matrix product density operators: Simulation of
  finite-temperature and dissipative systems},
\newblock Phys. Rev. Lett. \textbf{93}, 207204 (2004),
\newblock \doi{10.1103/PhysRevLett.93.207204}.

\bibitem{RevModPhys.91.021001}
D.~A. Abanin, E.~Altman, I.~Bloch and M.~Serbyn,
\newblock \emph{Colloquium: Many-body localization, thermalization, and
  entanglement},
\newblock Rev. Mod. Phys. \textbf{91}, 021001 (2019),
\newblock \doi{10.1103/RevModPhys.91.021001}.

\bibitem{NH_review}
R.~Nandkishore and D.~A. Huse,
\newblock \emph{{Many-Body Localization and Thermalization in Quantum
  Statistical Mechanics}},
\newblock Annual Review of Condensed Matter Physics \textbf{6}(1), 15 (2015),
\newblock \doi{10.1146/annurev-conmatphys-031214-014726},
\newblock \eprint{https://doi.org/10.1146/annurev-conmatphys-031214-014726}.

\bibitem{Bertini_review}
B.~Bertini, F.~Heidrich-Meisner, C.~Karrasch, T.~Prosen, R.~Steinigeweg and
  M.~Znidaric,
\newblock \emph{Finite-temperature transport in one-dimensional quantum lattice
  models},
\newblock arXiv preprint arXiv:2003.03334  (2020).

\bibitem{ETH_review}
{Luca D'Alessio and Yariv Kafri and Anatoli Polkovnikov and Marcos Rigol},
\newblock \emph{{From quantum chaos and eigenstate thermalization to
  statistical mechanics and thermodynamics}},
\newblock Advances in Physics \textbf{65}(3), 239 (2016),
\newblock \doi{10.1080/00018732.2016.1198134}.

\bibitem{Bloch08}
I.~Bloch, J.~Dalibard and W.~Zwerger,
\newblock \emph{{Many-body physics with ultracold gases}},
\newblock Reviews of Modern Physics \textbf{80}(3), 885 (2008),
\newblock \doi{10.1103/revmodphys.80.885}.

\bibitem{Bloch12}
I.~Bloch, J.~Dalibard and S.~Nascimb{\`{e}}ne,
\newblock \emph{{Quantum simulations with ultracold quantum gases}},
\newblock Nature Physics \textbf{8}(4), 267 (2012),
\newblock \doi{10.1038/nphys2259}.

\bibitem{Langen15}
T.~Langen, S.~Erne, R.~Geiger, B.~Rauer, T.~Schweigler, M.~Kuhnert,
  W.~Rohringer, I.~E. Mazets, T.~Gasenzer and J.~Schmiedmayer,
\newblock \emph{{Experimental observation of a generalized Gibbs ensemble}},
\newblock Science \textbf{348}(6231), 207 (2015),
\newblock \doi{10.1126/science.1257026}.

\bibitem{Schemmer19}
M.~Schemmer, I.~Bouchoule, B.~Doyon and J.~Dubail,
\newblock \emph{{Generalized Hydrodynamics on an Atom Chip}},
\newblock Physical Review Letters \textbf{122}(9) (2019),
\newblock \doi{10.1103/physrevlett.122.090601}.

\bibitem{Ketterle20}
P.~N. Jepsen, J.~Amato-Grill, I.~Dimitrova, W.~W. Ho, E.~Demler and
  W.~Ketterle,
\newblock \emph{{Spin transport in a tunable Heisenberg model realized with
  ultracold atoms}},
\newblock Nature \textbf{588}(7838), 403 (2020),
\newblock \doi{10.1038/s41586-020-3033-y}.

\bibitem{Malvania21}
Y.~Le, N.~Malvania, Y.~Zhang, J.~Dubail, M.~Rigol and D.~Weiss,
\newblock \emph{{Generalized hydrodynamics in strongly interacting 1D Bose
  gases (I)}},
\newblock Bulletin of the American Physical Society  (2021).

\bibitem{DDKY17}
B.~Doyon, J.~Dubail, R.~Konik and T.~Yoshimura,
\newblock \emph{{Large-Scale Description of Interacting One-Dimensional Bose
  Gases: Generalized Hydrodynamics Supersedes Conventional Hydrodynamics}},
\newblock Physical Review Letters \textbf{119}(19) (2017),
\newblock \doi{10.1103/physrevlett.119.195301}.

\bibitem{Misguich17}
G.~Misguich, K.~Mallick and P.~L. Krapivsky,
\newblock \emph{{Dynamics of the spin-$\frac{1}{2}$ Heisenberg chain
  initialized in a domain-wall state}},
\newblock Phys. Rev. B \textbf{96}, 195151 (2017),
\newblock \doi{10.1103/PhysRevB.96.195151}.

\bibitem{Ljubotina17-diff}
M.~Ljubotina, M.~{\v Z}nidari{\v c} and T.~Prosen,
\newblock \emph{A class of states supporting diffusive spin dynamics in the
  isotropic heisenberg model},
\newblock Journal of Physics A: Mathematical and Theoretical \textbf{50}(47),
  475002 (2017),
\newblock \doi{10.1088/1751-8121/aa8bdc}.

\bibitem{DYC18}
B.~Doyon, T.~Yoshimura and J.-S. Caux,
\newblock \emph{{Soliton Gases and Generalized Hydrodynamics}},
\newblock Physical Review Letters \textbf{120}(4) (2018),
\newblock \doi{10.1103/physrevlett.120.045301}.

\bibitem{SciPostPhys.2.2.014}
B.~Doyon and T.~Yoshimura,
\newblock \emph{{A note on generalized hydrodynamics: inhomogeneous fields and
  other concepts}},
\newblock SciPost Phys. \textbf{2}, 014 (2017),
\newblock \doi{10.21468/SciPostPhys.2.2.014}.

\bibitem{PhysRevLett.120.045301}
B.~Doyon, T.~Yoshimura and J.-S. Caux,
\newblock \emph{Soliton gases and generalized hydrodynamics},
\newblock Phys. Rev. Lett. \textbf{120}, 045301 (2018),
\newblock \doi{10.1103/PhysRevLett.120.045301}.

\bibitem{Bastianello_noise}
A.~Bastianello, J.~D. Nardis and A.~D. Luca,
\newblock \emph{{Generalized hydrodynamics with dephasing noise}},
\newblock Physical Review B \textbf{102}(16) (2020),
\newblock \doi{10.1103/physrevb.102.161110}.

\bibitem{CDDY19}
J.-S. Caux, B.~Doyon, J.~Dubail, R.~Konik and T.~Yoshimura,
\newblock \emph{{Hydrodynamics of the interacting Bose gas in the Quantum
  Newton Cradle setup}},
\newblock {SciPost} Physics \textbf{6}(6) (2019),
\newblock \doi{10.21468/scipostphys.6.6.070}.

\bibitem{BPK2019}
B.~Bertini, L.~Piroli and M.~Kormos,
\newblock \emph{{Transport in the sine-Gordon field theory: From generalized
  hydrodynamics to semiclassics}},
\newblock Physical Review B \textbf{100}(3) (2019),
\newblock \doi{10.1103/physrevb.100.035108}.

\bibitem{KlumperDrude}
A.~Urichuk, Y.~{\"O}z, A.~Kl{\"u}mper and J.~Sirker,
\newblock \emph{{The spin Drude weight of the XXZ chain and generalized
  hydrodynamics}},
\newblock SciPost Phys. \textbf{6}, 5 (2019),
\newblock \doi{10.21468/SciPostPhys.6.1.005}.

\bibitem{BPP20}
M.~Borsi, B.~Pozsgay and L.~Pristy{\'{a}}k,
\newblock \emph{{Current Operators in Bethe Ansatz and Generalized
  Hydrodynamics: An Exact Quantum-Classical Correspondence}},
\newblock Physical Review X \textbf{10}(1) (2020),
\newblock \doi{10.1103/physrevx.10.011054}.

\bibitem{CCR11}
A.~C. Cassidy, C.~W. Clark and M.~Rigol,
\newblock \emph{{Generalized Thermalization in an Integrable Lattice System}},
\newblock Physical Review Letters \textbf{106}(14) (2011),
\newblock \doi{10.1103/physrevlett.106.140405}.

\bibitem{VR_review}
L.~Vidmar and M.~Rigol,
\newblock \emph{{Generalized Gibbs ensemble in integrable lattice models}},
\newblock Journal of Statistical Mechanics: Theory and Experiment
  \textbf{2016}(6), 064007 (2016),
\newblock \doi{10.1088/1742-5468/2016/06/064007}.

\bibitem{PhysRevLett.115.157201}
E.~Ilievski, J.~{De Nardis}, B.~Wouters, J.-S. Caux, F.~H.~L. Essler and
  T.~Prosen,
\newblock \emph{{Complete Generalized Gibbs Ensembles in an Interacting
  Theory}},
\newblock Phys. Rev. Lett. \textbf{115}, 157201 (2015),
\newblock \doi{10.1103/PhysRevLett.115.157201}.

\bibitem{EF_review}
F.~H.~L. Essler and M.~Fagotti,
\newblock \emph{Quench dynamics and relaxation in isolated integrable quantum
  spin chains},
\newblock J. Stat. Mech. Theory Exp \textbf{2016}(6), 064002 (2016),
\newblock \doi{10.1088/1742-5468/2016/06/064002}.

\bibitem{BCDF16}
B.~Bertini, M.~Collura, J.~De~Nardis and M.~Fagotti,
\newblock \emph{{Transport in Out-of-Equilibrium $XXZ$ Chains: Exact Profiles
  of Charges and Currents}},
\newblock Phys. Rev. Lett. \textbf{117}, 207201 (2016),
\newblock \doi{10.1103/PhysRevLett.117.207201}.

\bibitem{CDY16}
O.~A. Castro-Alvaredo, B.~Doyon and T.~Yoshimura,
\newblock \emph{{Emergent Hydrodynamics in Integrable Quantum Systems Out of
  Equilibrium}},
\newblock Phys. Rev. X \textbf{6}, 041065 (2016),
\newblock \doi{10.1103/PhysRevX.6.041065}.

\bibitem{Bulchandani17}
V.~B. Bulchandani, R.~Vasseur, C.~Karrasch and J.~E. Moore,
\newblock \emph{{Solvable Hydrodynamics of Quantum Integrable Systems}},
\newblock Physical Review Letters \textbf{119}(22) (2017),
\newblock \doi{10.1103/physrevlett.119.220604}.

\bibitem{LjubotinaNature17}
M.~Ljubotina, M.~{\v{Z}}nidari{\v{c}} and T.~Prosen,
\newblock \emph{Spin diffusion from an inhomogeneous quench in an integrable
  system},
\newblock Nature Communications \textbf{8}(1) (2017),
\newblock \doi{10.1038/ncomms16117}.

\bibitem{Bulchandani18}
V.~B. Bulchandani, R.~Vasseur, C.~Karrasch and J.~E. Moore,
\newblock \emph{{Bethe-Boltzmann hydrodynamics and spin transport in the XXZ
  chain}},
\newblock Physical Review B \textbf{97}(4) (2018),
\newblock \doi{10.1103/physrevb.97.045407}.

\bibitem{LjubotinaPRL19}
M.~Ljubotina, M.~{\v{Z}}nidari{\v{c}} and T.~Prosen,
\newblock \emph{{Kardar-Parisi-Zhang Physics in the Quantum Heisenberg
  Magnet}},
\newblock Physical Review Letters \textbf{122}(21) (2019),
\newblock \doi{10.1103/physrevlett.122.210602}.

\bibitem{DupontMoore19}
M.~Dupont and J.~E. Moore,
\newblock \emph{Universal spin dynamics in infinite-temperature one-dimensional
  quantum magnets},
\newblock Physical Review B \textbf{101}(12) (2020),
\newblock \doi{10.1103/physrevb.101.121106}.

\bibitem{GV19}
S.~Gopalakrishnan and R.~Vasseur,
\newblock \emph{{Kinetic Theory of Spin Diffusion and Superdiffusion in XXZ
  Spin Chains}},
\newblock Physical Review Letters \textbf{122}(12) (2019),
\newblock \doi{10.1103/physrevlett.122.127202}.

\bibitem{GVW19}
S.~Gopalakrishnan, R.~Vasseur and B.~Ware,
\newblock \emph{{Anomalous relaxation and the high-temperature structure factor
  of XXZ spin chains}},
\newblock Proceedings of the National Academy of Sciences \textbf{116}(33),
  16250 (2019),
\newblock \doi{10.1073/pnas.1906914116}.

\bibitem{NMKI20}
J.~D. Nardis, M.~Medenjak, C.~Karrasch and E.~Ilievski,
\newblock \emph{{Universality Classes of Spin Transport in One-Dimensional
  Isotropic Magnets: The Onset of Logarithmic Anomalies}},
\newblock Physical Review Letters \textbf{124}(21) (2020),
\newblock \doi{10.1103/physrevlett.124.210605}.

\bibitem{Agrawal20}
U.~Agrawal, S.~Gopalakrishnan, R.~Vasseur and B.~Ware,
\newblock \emph{{Anomalous low-frequency conductivity in easy-plane XXZ spin
  chains}},
\newblock Phys. Rev. B \textbf{101}, 224415 (2020),
\newblock \doi{10.1103/PhysRevB.101.224415}.

\bibitem{GHDquenches_review}
V.~Alba, B.~Bertini, M.~Fagotti, L.~Piroli and P.~Ruggiero,
\newblock \emph{{Generalized-hydrodynamic approach to inhomogeneous quenches:
  Correlations, entanglement and quantum effects}},
\newblock arXiv preprint arXiv:2104.00656  (2021).

\bibitem{FF_review}
A.~C. Cubero, T.~Yoshimura and H.~Spohn,
\newblock \emph{{Form factors and generalized hydrodynamics for integrable
  systems}},
\newblock arXiv preprint arXiv:2104.04951  (2021).

\bibitem{diffusion_review}
J.~De~Nardis, B.~Doyon, M.~Medenjak and M.~Panfil,
\newblock \emph{Correlation functions and transport coefficients in generalised
  hydrodynamics},
\newblock arXiv preprint arXiv:2104.04462  (2021).

\bibitem{superdiffusion_review}
V.~B. Bulchandani, S.~Gopalakrishnan and E.~Ilievski,
\newblock \emph{Superdiffusion in spin chains},
\newblock arXiv preprint arXiv:2103.01976  (2021).

\bibitem{RCA54_review}
B.~Bu{\v{c}}a, K.~Klobas and T.~Prosen,
\newblock \emph{{Rule 54: Exactly solvable model of nonequilibrium statistical
  mechanics}},
\newblock arXiv preprint arXiv:2103.16543  (2021).

\bibitem{currents_review}
M.~Borsi, B.~Pozsgay and L.~Pristy{\'a}k,
\newblock \emph{{Current operators in integrable models: A review}},
\newblock arXiv preprint arXiv:2103.12160  (2021).

\bibitem{Bastianello18}
A.~Bastianello, B.~Doyon, G.~Watts and T.~Yoshimura,
\newblock \emph{{Generalized hydrodynamics of classical integrable field
  theory: the sinh-Gordon model}},
\newblock {SciPost} Physics \textbf{4}(6) (2018),
\newblock \doi{10.21468/scipostphys.4.6.045}.

\bibitem{DoyonToda19}
B.~Doyon,
\newblock \emph{{Generalized hydrodynamics of the classical Toda system}},
\newblock Journal of Mathematical Physics \textbf{60}(7), 073302 (2019),
\newblock \doi{10.1063/1.5096892}.

\bibitem{Gamayun19}
O.~Gamayun, Y.~Miao and E.~Ilievski,
\newblock \emph{{Domain-wall dynamics in the Landau-Lifshitz magnet and the
  classical-quantum correspondence for spin transport}},
\newblock Physical Review B \textbf{99}(14) (2019),
\newblock \doi{10.1103/physrevb.99.140301}.

\bibitem{SpohnToda20}
H.~Spohn,
\newblock \emph{{Collision rate ansatz for the classical Toda lattice}},
\newblock Physical Review E \textbf{101}(6) (2020),
\newblock \doi{10.1103/physreve.101.060103}.

\bibitem{SpohnJPA20}
H.~Spohn,
\newblock \emph{{Ballistic space-time correlators of the classical toda
  lattice}},
\newblock Journal of Physics A: Mathematical and Theoretical \textbf{53}(26),
  265004 (2020),
\newblock \doi{10.1088/1751-8121/ab91d5}.

\bibitem{Kuniba_2020}
A.~Kuniba, G.~Misguich and V.~Pasquier,
\newblock \emph{{Generalized hydrodynamics in box-ball system}},
\newblock Journal of Physics A: Mathematical and Theoretical \textbf{53}(40),
  404001 (2020),
\newblock \doi{10.1088/1751-8121/abadb9}.

\bibitem{Croydon_2020}
D.~A. Croydon and M.~Sasada,
\newblock \emph{{Generalized Hydrodynamic Limit for the Box{\textendash}Ball
  System}},
\newblock Communications in Mathematical Physics \textbf{383}(1), 427 (2020),
\newblock \doi{10.1007/s00220-020-03914-x}.

\bibitem{kuniba2020generalized}
A.~Kuniba, G.~Misguich and V.~Pasquier,
\newblock \emph{{Generalized hydrodynamics in complete box-ball system for
  $U_q(\widehat{sl}_n)$}},
\newblock arXiv preprint arXiv:2011.08052  (2020).

\bibitem{VanicatTrotterization}
M.~Vanicat, L.~Zadnik and T.~Prosen,
\newblock \emph{{Integrable Trotterization: Local Conservation Laws and
  Boundary Driving}},
\newblock Phys. Rev. Lett. \textbf{121}, 030606 (2018),
\newblock \doi{10.1103/PhysRevLett.121.030606}.

\bibitem{krajnik2020integrable}
{\v{Z}}.~Krajnik, E.~Ilievski and T.~Prosen,
\newblock \emph{Integrable matrix models in discrete space-time},
\newblock {SciPost} Physics \textbf{9}(3) (2020),
\newblock \doi{10.21468/scipostphys.9.3.038}.

\bibitem{Klobas18}
K.~Klobas, M.~Medenjak and T.~Prosen,
\newblock \emph{{Exactly solvable deterministic lattice model of crossover
  between ballistic and diffusive transport}},
\newblock Journal of Statistical Mechanics: Theory and Experiment
  \textbf{2018}(12), 123202 (2018),
\newblock \doi{10.1088/1742-5468/aae853}.

\bibitem{krajnik2020kardar}
{\v{Z}}.~Krajnik and T.~Prosen,
\newblock \emph{{Kardar{\textendash}Parisi{\textendash}Zhang Physics in
  Integrable Rotationally Symmetric Dynamics on Discrete Space{\textendash}Time
  Lattice}},
\newblock Journal of Statistical Physics \textbf{179}(1), 110 (2020),
\newblock \doi{10.1007/s10955-020-02523-1}.

\bibitem{GP17}
V.~Gritsev and A.~Polkovnikov,
\newblock \emph{{Integrable Floquet dynamics}},
\newblock {SciPost} Physics \textbf{2}(3) (2017),
\newblock \doi{10.21468/scipostphys.2.3.021}.

\bibitem{ljubotina2019ballistic}
M.~Ljubotina, L.~Zadnik and T.~Prosen,
\newblock \emph{{Ballistic Spin Transport in a Periodically Driven Integrable
  Quantum System}},
\newblock Physical Review Letters \textbf{122}(15) (2019),
\newblock \doi{10.1103/physrevlett.122.150605}.

\bibitem{Takhtajan77}
L.~Takhtajan,
\newblock \emph{{Integration of the continuous Heisenberg spin chain through
  the inverse scattering method}},
\newblock Physics Letters A \textbf{64}(2), 235 (1977),
\newblock \doi{10.1016/0375-9601(77)90727-7}.

\bibitem{Sklyanin79}
E.~Sklyanin,
\newblock \emph{{On complete integrability of the Landau-Lifshitz equation}}
  (1979).

\bibitem{Faddeev1987}
L.~D. Faddeev and L.~A. Takhtajan,
\newblock \emph{{Hamiltonian Methods in the Theory of Solitons}},
\newblock Springer Berlin Heidelberg,
\newblock \doi{10.1007/978-3-540-69969-9} (1987).

\bibitem{Hirota82}
R.~Hirota,
\newblock \emph{{Bilinearization of Soliton Equations}},
\newblock Journal of the Physical Society of Japan \textbf{51}(1), 323 (1982),
\newblock \doi{10.1143/jpsj.51.323}.

\bibitem{Lax1968}
P.~D. Lax,
\newblock \emph{{Integrals of nonlinear equations of evolution and solitary
  waves}},
\newblock Communications on Pure and Applied Mathematics \textbf{21}(5), 467
  (1968),
\newblock \doi{10.1002/cpa.3160210503}.

\bibitem{AKNS1974}
M.~J. Ablowitz, D.~J. Kaup, A.~C. Newell and H.~Segur,
\newblock \emph{{The Inverse Scattering Transform-Fourier Analysis for
  Nonlinear Problems}},
\newblock Studies in Applied Mathematics \textbf{53}(4), 249 (1974),
\newblock \doi{10.1002/sapm1974534249}.

\bibitem{AblowitzSegur_book}
M.~J. Ablowitz and H.~Segur,
\newblock \emph{{Solitons and the Inverse Scattering Transform}},
\newblock Society for Industrial and Applied Mathematics,
\newblock \doi{10.1137/1.9781611970883} (1981).

\bibitem{Babelon_book}
O.~Babelon, D.~Bernard and M.~Talon,
\newblock \emph{{Introduction to Classical Integrable Systems}},
\newblock Cambridge University Press,
\newblock \doi{10.1017/cbo9780511535024} (2003).

\bibitem{Hietarinta_book}
J.~Hietarinta, N.~Joshi and F.~W. Nijhoff,
\newblock \emph{{Discrete Systems and Integrability}},
\newblock Cambridge University Press,
\newblock \doi{10.1017/cbo9781107337411} (2016).

\bibitem{Moser_1991}
J.~Moser and A.~P. Veselov,
\newblock \emph{{Discrete versions of some classical integrable systems and
  factorization of matrix polynomials}},
\newblock Communications in Mathematical Physics \textbf{139}(2), 217 (1991),
\newblock \doi{10.1007/bf02352494}.

\bibitem{Veselov_2003}
A.~Veselov,
\newblock \emph{{Yang{\textendash}Baxter maps and integrable dynamics}},
\newblock Physics Letters A \textbf{314}(3), 214 (2003),
\newblock \doi{10.1016/s0375-9601(03)00915-0}.

\bibitem{Sklyanin83}
E.~K. Sklyanin,
\newblock \emph{{Some algebraic structures connected with the Yang--Baxter
  equation}},
\newblock Functional Analysis and Its Applications \textbf{16}(4), 263 (1983),
\newblock \doi{10.1007/bf01077848}.

\bibitem{PZ13}
T.~Prosen and B.~{\v Z}unkovi{\v c},
\newblock \emph{{Macroscopic Diffusive Transport in a Microscopically
  Integrable Hamiltonian System}},
\newblock Phys. Rev. Lett. \textbf{111}, 040602 (2013),
\newblock \doi{10.1103/PhysRevLett.111.040602}.

\bibitem{DasKPZ}
A.~Das, M.~Kulkarni, H.~Spohn and A.~Dhar,
\newblock \emph{{Kardar-Parisi-Zhang scaling for an integrable lattice
  Landau-Lifshitz spin chain}},
\newblock Phys. Rev. E \textbf{100}, 042116 (2019),
\newblock \doi{10.1103/PhysRevE.100.042116}.

\bibitem{Sklyanin89XYZ}
E.~K. Sklyanin,
\newblock \emph{{Poisson structure of a periodic classical XYZ-chain}},
\newblock Journal of Soviet Mathematics \textbf{46}(1), 1664 (1989),
\newblock \doi{10.1007/bf01099198}.

\bibitem{FRT88}
L.~Faddeev, N.~Reshetikhin and L.~Takhtajan,
\newblock \emph{{Quantization of Lie Groups and Lie Algebras}},
\newblock In \emph{Algebraic Analysis}, pp. 129--139. Elsevier,
\newblock \doi{10.1016/b978-0-12-400465-8.50019-5} (1988).

\bibitem{Drinfeld88}
V.~G. Drinfeld,
\newblock \emph{Quantum groups},
\newblock Journal of Soviet Mathematics \textbf{41}(2), 898 (1988),
\newblock \doi{10.1007/bf01247086}.

\bibitem{KT96}
S.~M. Khoroshkin and V.~N. Tolstoy,
\newblock \emph{Yangian double},
\newblock Letters in Mathematical Physics \textbf{36}(4), 373 (1996),
\newblock \doi{10.1007/bf00714404}.

\bibitem{KT14}
S.~Khoroshkin and Z.~Tsuboi,
\newblock \emph{{The universal R-matrix and factorization of the L-operators
  related to the Baxter Q-operators}},
\newblock Journal of Physics A: Mathematical and Theoretical \textbf{47}(19),
  192003 (2014),
\newblock \doi{10.1088/1751-8113/47/19/192003}.

\bibitem{Baxter_book}
R.~J. Baxter,
\newblock \emph{{Exactly Solved Models in Statistical Mechanics}},
\newblock In \emph{Series on Advances in Statistical Mechanics}, pp. 5--63.
  {WORLD} {SCIENTIFIC},
\newblock \doi{10.1142/9789814415255_0002} (1985).

\bibitem{DerkachovI}
S.~E. Derkachov,
\newblock \emph{{Factorization of the R-matrix. I}},
\newblock Journal of Mathematical Sciences \textbf{143}(1), 2773 (2007),
\newblock \doi{10.1007/s10958-007-0164-8}.

\bibitem{DerkachovII}
S.~E. Derkachov,
\newblock \emph{{Factorization of the R-matrix. II}},
\newblock Journal of Mathematical Sciences \textbf{143}(1), 2791 (2007),
\newblock \doi{10.1007/s10958-007-0165-7}.

\bibitem{Derkachov06}
S.~Derkachov, D.~Karakhanyan and R.~Kirschner,
\newblock \emph{{Baxter Q-operators of the XXZ chain and R-matrix
  factorization}},
\newblock Nuclear Physics B \textbf{738}(3), 368 (2006),
\newblock \doi{10.1016/j.nuclphysb.2005.12.015}.

\bibitem{Bazhanov10}
V.~V. Bazhanov, T.~{\L}ukowski, C.~Meneghelli and M.~Staudacher,
\newblock \emph{{A shortcut to the Q-operator}},
\newblock Journal of Statistical Mechanics: Theory and Experiment
  \textbf{2010}(11), P11002 (2010),
\newblock \doi{10.1088/1742-5468/2010/11/p11002}.

\bibitem{Frassek11}
R.~Frassek, T.~{\L}ukowski, C.~Meneghelli and M.~Staudacher,
\newblock \emph{{Oscillator construction of Q-operators}},
\newblock Nuclear Physics B \textbf{850}(1), 175 (2011),
\newblock \doi{10.1016/j.nuclphysb.2011.04.008}.

\bibitem{Babelon_2018}
O.~Babelon, K.~K. Kozlowski and V.~Pasquier,
\newblock \emph{{Baxter Operator and Baxter Equation for q-Toda and Toda2
  Chains}},
\newblock Reviews in Mathematical Physics \textbf{30}(06), 1840003 (2018),
\newblock \doi{10.1142/s0129055x18400032}.

\bibitem{Veselov03}
A.~Veselov,
\newblock \emph{{Yang{\textendash}Baxter maps and integrable dynamics}},
\newblock Physics Letters A \textbf{314}(3), 214 (2003),
\newblock \doi{10.1016/s0375-9601(03)00915-0}.

\bibitem{BS18}
V.~V. Bazhanov and S.~M. Sergeev,
\newblock \emph{{Yang{\textendash}Baxter maps, discrete integrable equations
  and quantum groups}},
\newblock Nuclear Physics B \textbf{926}, 509 (2018),
\newblock \doi{10.1016/j.nuclphysb.2017.11.017}.

\bibitem{Tsuboi18}
Z.~Tsuboi,
\newblock \emph{{Quantum groups, Yang{\textendash}Baxter maps and
  quasi-determinants}},
\newblock Nuclear Physics B \textbf{926}, 200 (2018),
\newblock \doi{10.1016/j.nuclphysb.2017.11.005}.

\bibitem{Doyon_lectures}
B.~Doyon,
\newblock \emph{{Lecture notes on Generalised Hydrodynamics}},
\newblock {SciPost} Physics Lecture Notes  (2020),
\newblock \doi{10.21468/scipostphyslectnotes.18}.

\bibitem{MarkoKPZ}
M.~{\v Z}nidari{\v c},
\newblock \emph{{Spin Transport in a One-Dimensional Anisotropic Heisenberg
  Model}},
\newblock Phys. Rev. Lett. \textbf{106}, 220601 (2011),
\newblock \doi{10.1103/PhysRevLett.106.220601}.

\bibitem{Ilievski18}
E.~Ilievski, J.~De~Nardis, M.~Medenjak and T.~Prosen,
\newblock \emph{{Superdiffusion in One-Dimensional Quantum Lattice Models}},
\newblock Phys. Rev. Lett. \textbf{121}, 230602 (2018),
\newblock \doi{10.1103/PhysRevLett.121.230602}.

\bibitem{NMKI19}
J.~De~Nardis, M.~Medenjak, C.~Karrasch and E.~Ilievski,
\newblock \emph{{Anomalous Spin Diffusion in One-Dimensional
  Antiferromagnets}},
\newblock Phys. Rev. Lett. \textbf{123}, 186601 (2019),
\newblock \doi{10.1103/PhysRevLett.123.186601}.

\bibitem{Vir20}
V.~B. Bulchandani,
\newblock \emph{{Kardar-Parisi-Zhang universality from soft gauge modes}},
\newblock Physical Review B \textbf{101}(4) (2020),
\newblock \doi{10.1103/physrevb.101.041411}.

\bibitem{DGIV20}
J.~De~Nardis, S.~Gopalakrishnan, E.~Ilievski and R.~Vasseur,
\newblock \emph{{Superdiffusion from Emergent Classical Solitons in Quantum
  Spin Chains}},
\newblock Phys. Rev. Lett. \textbf{125}, 070601 (2020),
\newblock \doi{10.1103/PhysRevLett.125.070601}.

\bibitem{superuniversality}
E.~Ilievski, J.~De~Nardis, S.~Gopalakrishnan, R.~Vasseur and B.~Ware,
\newblock \emph{{Superuniversality of superdiffusion}},
\newblock arXiv preprint arXiv:2009.08425  (2020).

\bibitem{KPZ86}
M.~Kardar, G.~Parisi and Y.-C. Zhang,
\newblock \emph{{Dynamic Scaling of Growing Interfaces}},
\newblock Phys. Rev. Lett. \textbf{56}, 889 (1986),
\newblock \doi{10.1103/PhysRevLett.56.889}.

\bibitem{IN_Drude}
E.~Ilievski and J.~De~Nardis,
\newblock \emph{{Microscopic Origin of Ideal Conductivity in Integrable Quantum
  Models}},
\newblock Phys. Rev. Lett. \textbf{119}, 020602 (2017),
\newblock \doi{10.1103/PhysRevLett.119.020602}.

\bibitem{Mazur69}
P.~Mazur,
\newblock \emph{Non-ergodicity of phase functions in certain systems},
\newblock Physica \textbf{43}(4), 533 (1969),
\newblock \doi{10.1016/0031-8914(69)90185-2}.

\bibitem{Suzuki71}
M.~Suzuki,
\newblock \emph{{Ergodicity, constants of motion, and bounds for
  susceptibilities}},
\newblock Physica \textbf{51}(2), 277 (1971),
\newblock \doi{10.1016/0031-8914(71)90226-6}.

\bibitem{Ilievski12}
E.~Ilievski and T.~Prosen,
\newblock \emph{{Thermodyamic Bounds on Drude Weights in Terms of
  Almost-conserved Quantities}},
\newblock Communications in Mathematical Physics \textbf{318}(3), 809 (2012),
\newblock \doi{10.1007/s00220-012-1599-4}.

\bibitem{Doyon2019_projections}
B.~Doyon,
\newblock \emph{Diffusion and superdiffusion from hydrodynamic projection},
\newblock arXiv preprint arXiv:1912.01551  (2019).

\bibitem{IMP15}
E.~Ilievski, M.~Medenjak and T.~Prosen,
\newblock \emph{{Quasilocal Conserved Operators in the Isotropic Heisenberg
  Spin-1/2 Chain}},
\newblock Physical Review Letters \textbf{115}(12) (2015),
\newblock \doi{10.1103/physrevlett.115.120601}.

\bibitem{quasilocal_review}
E.~Ilievski, M.~Medenjak, T.~Prosen and L.~Zadnik,
\newblock \emph{Quasilocal charges in integrable lattice systems},
\newblock Journal of Statistical Mechanics: Theory and Experiment
  \textbf{2016}(6), 064008 (2016),
\newblock \doi{10.1088/1742-5468/2016/06/064008}.

\bibitem{Prosen11}
T.~Prosen,
\newblock \emph{{Open $XXZ$ Spin Chain: Nonequilibrium Steady State and a
  Strict Bound on Ballistic Transport}},
\newblock Phys. Rev. Lett. \textbf{106}, 217206 (2011),
\newblock \doi{10.1103/PhysRevLett.106.217206}.

\bibitem{PI13}
T.~Prosen and E.~Ilievski,
\newblock \emph{{Families of Quasilocal Conservation Laws and Quantum Spin
  Transport}},
\newblock Phys. Rev. Lett. \textbf{111}, 057203 (2013),
\newblock \doi{10.1103/PhysRevLett.111.057203}.

\bibitem{ProsenNPB14}
T.~Prosen,
\newblock \emph{{Quasilocal conservation laws in XXZ spin-1/2 chains: Open,
  periodic and twisted boundary conditions}},
\newblock Nuclear Physics B \textbf{886}, 1177 (2014),
\newblock \doi{10.1016/j.nuclphysb.2014.07.024}.

\bibitem{Pereira14}
R.~G. Pereira, V.~Pasquier, J.~Sirker and I.~Affleck,
\newblock \emph{{Exactly conserved quasilocal operators for the XXZ spin
  chain}},
\newblock Journal of Statistical Mechanics: Theory and Experiment
  \textbf{2014}(9), P09037 (2014),
\newblock \doi{10.1088/1742-5468/2014/09/p09037}.

\bibitem{DS17}
B.~Doyon and H.~Spohn,
\newblock \emph{{Drude Weight for the Lieb-Liniger Bose Gas}},
\newblock {SciPost} Physics \textbf{3}(6) (2017),
\newblock \doi{10.21468/scipostphys.3.6.039}.

\bibitem{IN_Hubbard}
E.~Ilievski and J.~D. Nardis,
\newblock \emph{{Ballistic transport in the one-dimensional Hubbard model: The
  hydrodynamic approach}},
\newblock Physical Review B \textbf{96}(8) (2017),
\newblock \doi{10.1103/physrevb.96.081118}.

\bibitem{DDB18}
J.~D. Nardis, D.~Bernard and B.~Doyon,
\newblock \emph{{Hydrodynamic Diffusion in Integrable Systems}},
\newblock Physical Review Letters \textbf{121}(16) (2018),
\newblock \doi{10.1103/physrevlett.121.160603}.

\bibitem{Gopalakrishnan18}
S.~Gopalakrishnan, D.~A. Huse, V.~Khemani and R.~Vasseur,
\newblock \emph{{Hydrodynamics of operator spreading and quasiparticle
  diffusion in interacting integrable systems}},
\newblock Physical Review B \textbf{98}(22) (2018),
\newblock \doi{10.1103/physrevb.98.220303}.

\bibitem{DDB19}
J.~D. Nardis, D.~Bernard and B.~Doyon,
\newblock \emph{{Diffusion in generalized hydrodynamics and quasiparticle
  scattering}},
\newblock {SciPost} Physics \textbf{6}(4) (2019),
\newblock \doi{10.21468/scipostphys.6.4.049}.

\bibitem{Novikov_book}
S.~Novikov, S.~Manakov, L.~Pitaevskii and V.~E. Zakharov,
\newblock \emph{Theory of solitons: the inverse scattering method},
\newblock Springer Science \& Business Media,
\newblock \doi{10.1007/978-3-642-81448-8_7} (1984).

\bibitem{Korepin_book}
V.~E. Korepin, N.~M. Bogoliubov and A.~G. Izergin,
\newblock \emph{{Quantum Inverse Scattering Method and Correlation Functions}},
\newblock Cambridge Monographs on Mathematical Physics. Cambridge University
  Press,
\newblock \doi{10.1017/CBO9780511628832} (1993).

\bibitem{Sklyanin88}
E.~K. Sklyanin,
\newblock \emph{{Classical limits of the SU(2)-invariant solutions of the
  Yang-Baxter equation}},
\newblock Journal of Soviet Mathematics \textbf{40}(1), 93 (1988),
\newblock \doi{10.1007/bf01084941}.

\bibitem{Sklyanin89XXZ}
E.~K. Sklyanin,
\newblock \emph{{Poisson structure of classical XXZ-chain}},
\newblock Journal of Soviet Mathematics \textbf{46}(5), 2104 (1989),
\newblock \doi{10.1007/bf01096094}.

\bibitem{Ishimori82}
Y.~Ishimori,
\newblock \emph{{An Integrable Classical Spin Chain}},
\newblock Journal of the Physical Society of Japan \textbf{51}(11), 3417
  (1982),
\newblock \doi{10.1143/jpsj.51.3417}.

\end{thebibliography}

\end{document}